\documentclass[journal]{IEEEtran}
\usepackage{times}
\usepackage{amsmath,amsfonts,amssymb}
\usepackage{algorithm}
\usepackage{algorithmic}
\usepackage{xcolor}

\providecommand{\vct}[1]{{\pmb #1}}       

    \providecommand{\abs}[1]{\lvert {#1} \rvert}                

    \DeclareMathOperator{\E}{E}                         

\providecommand{\norm}[1]{\lVert {#1} \rVert}

\newcommand{\eqnum}[2][noname]{\begin{equation}\label{#1} \begin{split} #2 \end{split} \end{equation} }

\usepackage{graphicx}
\usepackage{multirow}
\usepackage{array}
\usepackage{booktabs}
\usepackage{subfigure} 
\usepackage{float}
\usepackage{fancyhdr}
\usepackage{ulem}

\graphicspath{{eps/}{tiff/}}
\graphicspath{{pic/}}
\graphicspath{{usedpic/}}
\usepackage{epstopdf}

\providecommand{\norm}[1]{\lVert {#1} \rVert}

\ifCLASSINFOpdf
\else
\fi
\begin{document}
%
\title{Carrier Phase Ranging for Indoor Positioning with 5G NR Signals}
%
%
%

\author{Liang~Chen,
	    Xin~Zhou,
	    Feifei~Chen,
        Lie-Liang Yang,~\IEEEmembership{Fellow,~IEEE,}
        and~Ruizhi~Chen
\thanks{Liang Chen, Xin Zhou, Feifei Chen and Ruizhi Chen are with the State Key Laboratory of Information Engineering in Surveying, Mapping and Remote Sensing, Wuhan University, Wuhan	430072, China (e-mail: l.chen@whu.edu.cn; x.zhou.whu@whu.edu.cn; cff5452@whu.edu.cn; ruizhi.chen@whu.edu.cn).
	
	Lie-Liang Yang is with the faculty of Electronics and Computer Science, University of Southampton, University Road, Southampton, SO17 1BJ, United Kingdom (email: lly@ecs.soton.ac.uk). 
}
}

%
%


\markboth{}%
{Shell \MakeLowercase{\textit{et al.}}: Bare Demo of IEEEtran.cls for IEEE Journals}

\pagestyle{fancy}


%



\maketitle

\begin{abstract}
Indoor positioning is one of the core technologies of Internet of Things (IoT) and artificial intelligence (AI), and is expected to play a significant role in the upcoming era of AI. However, affected by the complexity of indoor environments, it is still highly challenging to achieve continuous and reliable indoor positioning. Currently, 5G  cellular networks are being deployed worldwide, the new technologies of which have brought the approaches for improving the performance of wireless indoor positioning. In this paper, we investigate the indoor positioning under the 5G new radio (NR), which has been standardized and being commercially operated in massive markets. Specifically, a solution is proposed and a software defined receiver (SDR) is developed for indoor positioning. With our SDR indoor positioning system, the 5G NR signals are firstly sampled by universal software radio peripheral (USRP), and then, coarse synchronization is achieved via detecting the start of the synchronization signal block (SSB). Then, with the assistance of the pilots transmitted on the physical broadcasting channel (PBCH), multipath acquisition and delay tracking are sequentially carried out to estimate the time of arrival (ToA) of received signals. Furthermore, to improve the ToA ranging accuracy, the carrier phase of the first arrived path is estimated. Finally, to quantify the accuracy of our ToA estimation method, indoor field tests are carried out in an office environment, where a 5G NR base station (known as gNB) is installed for commercial use. Our test results show that, in the static test scenarios, the ToA accuracy measured by the 1-$\sigma$ error interval is about 0.5 m, while in the pedestrian mobile environment, the probability of range accuracy within 0.8 m is 95\%. 
\end{abstract}


%
\IEEEpeerreviewmaketitle

\thispagestyle{fancy}
\setlength\headheight{8pt}
\lhead{\scriptsize \qquad This article has been accepted for publication in IEEE Internet of Things Journal, but has not been fully edited. Content may change prior to final publication. \\ \textbf{Citation information:} L. Chen, X. Zhou, F. Chen, L. -L. Yang and R. Chen, "Carrier Phase Ranging for Indoor Positioning with 5G NR Signals," IEEE Internet of Things\\
\qquad \qquad \qquad \qquad \qquad \qquad \qquad  \qquad \qquad \qquad \qquad \qquad Journal, Nov. 2021, doi: 10.1109/JIOT.2021.3125373} 
\renewcommand\headrulewidth{0pt} 

\section{Introduction}

%
%
%
%
\IEEEPARstart{W}{ith} the extensive urban development, indoor positioning is becoming more and more important. According to the report of the U.S. Environmental Protection Agency, people spend nearly 70\%--90\% of their time indoors~\cite{cite01}. Indoor positioning plays fundamental roles in a wide area of Internet of Things (IoT) based applications, such as indoor emergent rescue~\cite{cite02}, precision marketing in shopping malls, asset manager and tracking in smart factory, mobile health services, virtual reality games, and location based social media, etc.~\cite{cite03}--\cite{cite04}. According to Market\&Markets research data,  the compound growth rate of global market size related to indoor positioning is about 22.5\% annually and it is projected to reach 18.74 billion USD by 2025~\cite{cite01},~\cite{cite02}. 

Despite the growing demand in the massive market, it is not an easy task to provide feasible indoor positioning solutions in many cases. As one of the most popular positioning technology, Global Navigation Satellite Systems (GNSS) has achieved great success in positioning in outdoor open area, and the positioning accuracy is able to reach sub-meter level when enhanced by various technologies ~\cite{cite04}. However, due to the low  power, GNSS signals cannot be well received indoor to provide continuous and reliable positioning. In many cases, especially in deep indoor areas, GNSS signals may be totally blocked. 

Indoor positioning has been extensively studied and various technologies have been developed based on, for example, WiFi, Bluetooth, ultra-wideband (UWB), pseudolites, geomagnetism, sound/ultrasound, or pedestrian dead reckoning (PDR), etc.\cite{cite04}. Although different technologies have their unique advantages, it is still highly challenging to achieve the accurate, effective, reliable and real-time positioning solutions for indoor applications due to the constraints of complex spatial layout, topological structure, and sophisticated  environment for signal transmission indoors~\cite{cite04}.

At present, the 5-th generation (5G) mobile networks  have been developed rapidly. In 2019, 5G cellular networks started to be deployed worldwide. According to the Ericsson Mobility Report, there are about 220 million 5G users worldwide by the end of 2020~\cite{cite06}. In China, 5G NR (new radio) was put into commercial operation in October 2019, and the 5G users exceeded 50 million in March 2020, which account for about 70 percent of the total number in the world~\cite{cite07}. As known from literature, 5G has introduced a range of technologies, including large-scale antenna array, ultra-dense networking, new multi-access schemes, full-spectrum access and the new network architecture based on software-defined networks (SDN), etc., in order to support the new-generation wireless communication with the characteristics of high capacity, high reliability and low delay in wide coverage areas~\cite{cite08}.

It has also been envisioned that these innovative technologies of 5G are beneficial to wireless positioning, which have drawn a lot of researches on the positioning based on 5G NR signals.  In general, these methods can be classified into two main categories, namely the geometry-based methods and feature matching-based methods, as further discussed below:

The geometry-based methods can be further divided into triangulation methods~\cite{cite08-AOA-1}--\cite{cite08-AOA-6}, trilateration methods~\cite{cite08-TOA-1}--\cite{cite08-TOA-4} and the joint estimation methods~\cite{cite08-Joint-1}--\cite{cite08-Joint-5}. 
Specifically, for the triangulation methods, the angle of arrival (AoA) and angle of departure (AoD) can be estimated by applying the techniques of massive MIMO (multiple input multiple output) and smart antennas in 5G. Correspondingly, different estimation methods have been proposed, which  include the classical MUSIC~\cite{cite08-AOA-1}--\cite{cite08-AOA-2}, ESPRIT~\cite{cite08-AOA-3}, combined far-field and near-field direct positioning (CNFDP)~\cite{cite08-AOA-4}, distributed compressed sensing-simultaneous orthogonal matching pursuit (DCS-SOMP)~\cite{cite08-AOA-5}, and the two-stage extended Kalman filter (EKF) based positioning method~\cite{cite08-AOA-6}. 
In the context of the trilateration methods, different timing-based methods have been proposed and investigated. For example, in~\cite{cite08-TOA-1}, the Primary Synchronization Signal (PSS) and Secondary Synchronization Signal (SSS) in the 5G standard were studied to obtain the Observed Time Difference of Arrival (OTDoA) estimation, where the performance of OTDoA and the impacts from different factors were comprehensively analyzed by simulations. In~\cite{cite08-TOA-2}, the NRPos Algorithm (NRPA) was proposed by using the timing advance (TA) information to determine user locations, and simulation results show that in urban/dense environments, the mean positioning accuracy of NRPA is 30-40\% better than that of the OTDoA method. In~\cite{cite08-TOA-3},~\cite{cite08-TOA-4}, the Positioning Reference Signals (PRS) was used to obtain the measurements of Time Difference of Arrival (TDoA), showing that the achievable positioning accuracy is within 1 meter.
Furthermore, in the joint estimation methods designed based on angle and timing, the 5G uplink reference signals are exploited to estimate and track the AoAs and ToAs. As some examples, a cascaded EKF was proposed and studied for positioning and tracking, which is capable of achieving the accuracy within sub-meter~\cite{cite08-Joint-1}. In~\cite{cite08-Joint-2}, the tensor-ESPRIT was proposed to jointly estimate AoD, AoA and ToA embedded in the geometric channel, and the maximum likelihood (ML) was applied for simultaneously positioning and mapping. Considering the downlink transmission of a multiple-input single-output (MISO) millimeter-wave (mmWave) system, TDoA and AoD are jointly estimated to estimate the position of a mobile station (MS)~\cite{cite08-Joint-3}, which achieve positioning accuracy of sub-meter.  
The positioning method based on ToA and AoA estimation by exploiting the co-band 5G signals is proposed in~\cite{cite08-Joint-4}, can also attain the positioning accuracy within sub-meter.
Recently, the authors of~\cite{cite08-Joint-5} proposed to use the mmWave PRS to obtain the AoA and ToA estimation,  and where the constraint satisfaction problem (CSP) is further formulated to improve the positioning accuracy in the 5G scenario with the aid of cooperative localization.

The feature matching-based methods are mainly referred to as the fingerprinting methods, which have also received wide attention in 5G positioning. Specifically, in 5G positioning, the received signal strength indicator (RSSI)-based fingerprinting and Channel State Information (CSI)-based fingerprinting have been investigated~\cite{cite08-FP-1}--\cite{cite08-CSI-2}. 
In RSSI based fingerprinting,  the Reference Signals Received Power (RSRP) is commonly utilized to quantify the received signal strength, which is derived from the measurement of the DL-RS (downlink reference signal)~\cite{cite08-FP-1}--\cite{cite08-FP-2}. By applying the different feature matching methods, such as K-Nearest Neighbors (KNN)~\cite{cite08-FP-1}, Neural Network (NN)~\cite{cite08-FP-2} or Random Forest (RF)~\cite{cite08-FP-2}, the positioning accuracy can be within 10 meters outdoor. In~\cite{cite08-FP-3}, the measurements of beam RSRP were used and with the aid of a two-layer Deep Neural Network (DNN), the mean positioning error is about 1.4 m. Based on the beamforming fingerprints, the Convolutional Neural Network (CNN) assisted approaches can achieve the average positioning error of about 1.78 m~\cite{cite08-FP-4} or 3.3 m~\cite{cite08-FP-5}, respectively, in different outdoor scenarios. In~\cite{cite08-FP-6}, by exploiting the Synchronization Signal (SS) RSRP, the Deep Learning (DL) method is capable of achieving the positioning accuracy of meter-level.  
Comparatively, we found that the studies relied on the 5G CSI-based fingerprinting methods relatively rare. On this topic, the authors of~\cite{cite08-CSI-1} exploited the angle-based measurements and proposed the Binning-based method, which can achieve the positioning accuracy slightly  below 10 meters outdoor. By contrast, the authors of~\cite{cite08-CSI-2} studied the uplink CSI-based localization and the achieved positioning accuracy is about 10 meters.

Recently, the importance of the Location Based Services (LBS) in 5G networks has been emphasised by 3GPP, which has the capability to enhance 5G positioning, as finalized in 3GPP Rel-16. To this purpose, the reference signals specifically introduced for positioning were included to the NR specifications. These include the PRS in the downlink and the sounding reference signal (SRS) in the uplink. Based on these signals, the 3GPP Rel-16 has introduced the advanced positioning solutions suitable for 5G NR, including the  angle-based positioning solution relied on the  downlink angle of departure (DL-AoD) or uplink angle of arrival (UL-AoA), and the time-based positioning solution designed on the downlink time difference of arrival (DL-TDoA), the uplink time difference of arrival (UL-TDoA), and Multi-cell round trip time (Multi-RTT).

In addition, with the introduction of Ultra Dense Network (UDN) in 5G, cooperative positioning and random geometry method have been shown to be able to achieve high accuracy positioning~\cite{cite08-AOA-6},~\cite{cite08-TOA-1},~\cite{cite08-TOA-3},~\cite{cite08-Joint-1}. Moreover, the integration of 5G signals with the signals obtained from other sensors designed specifically for positioning have been investigated the context of wireless positioning in~\cite{cite08-fusion-1}--\cite{cite08-fusion-3}.

Although there are a lot of researches effort on the 5G positioning, there are still numerous challenges. First, it is well known that the feature matching-based methods is time consuming to build up the database of fingerprints and hard to update, especially, in the dynamic environments, which hence has the limit for wide deployment in practice. Second, the geometry-based method in mmWave positioning is a promising method. However, mmWave communication has so far not been commercially implemented in 5G and the specification has even not been well defined by 3GPP. 
Finally, for the timing-based positioning, the researches have mainly been carried out via numerical simulations or laboratory emulation of 5G signals, while the investigation in practical scenarios, especially for indoors, is very limited~\cite{cite08-NR}. 

\begin{figure}[t]
	\centering
	\includegraphics [width=.8\linewidth]{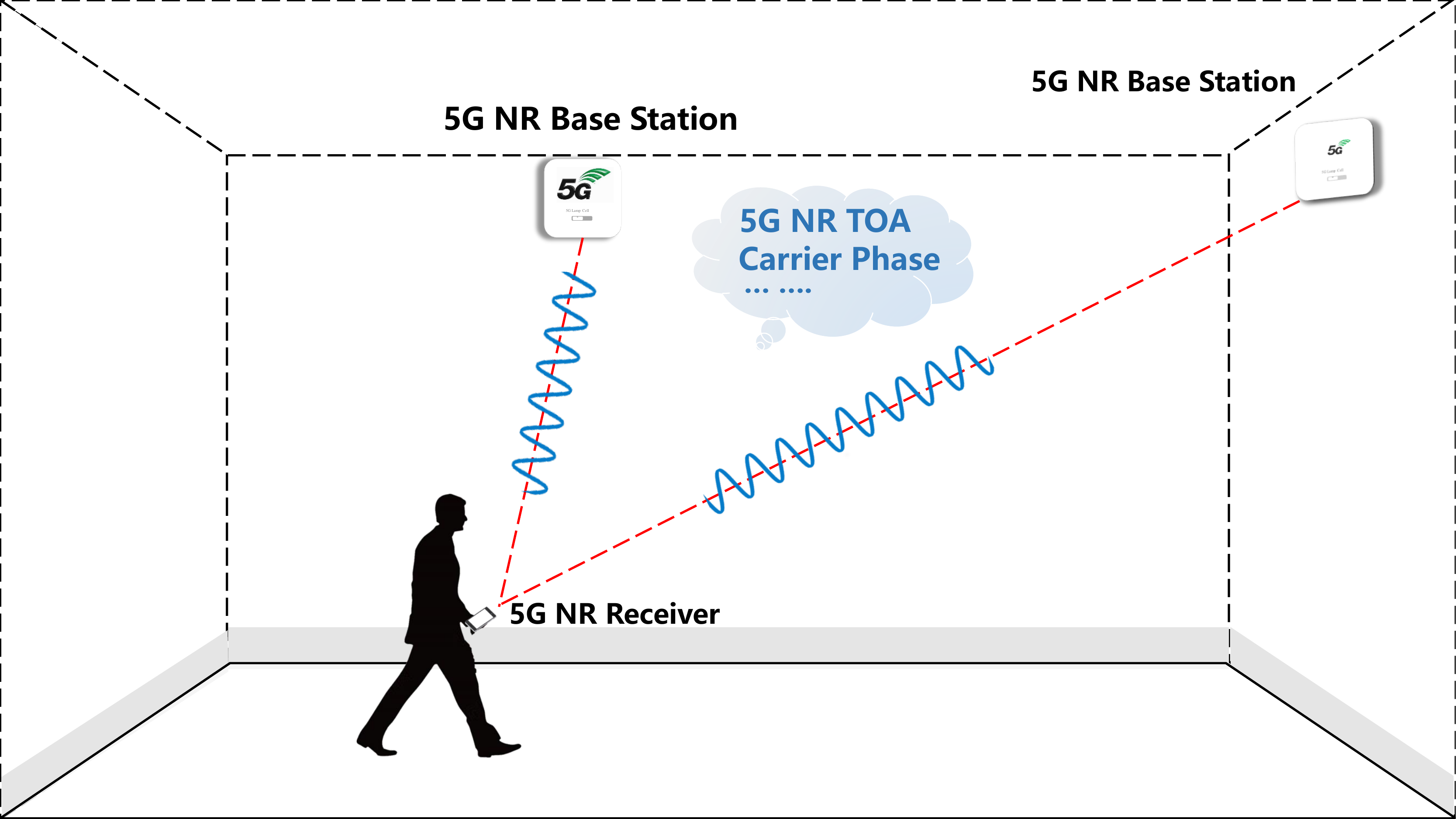}\\
	\caption{Carrier phase ranging with 5G NR signals for indoor positioning.}
	\label{fig:methodframe}
\end{figure}

Against the background, in this paper, we aim at developing a practical positioning method for massive smartphone users to  estimate the positioning parameters and compute the locations of their own. Thus, our method is a user-centric positioning method. Currently, the operational frequency of 5G NR for commercial use is sub-6 GHz. Constrained by the limited size of smartphones, antenna arrays are difficult to be installed. As a consequence, the angle based positioning methods cannot be implemented. Therefore, in this paper, we resort to the timing-based positioning method via exploiting the demodulation reference signal (DM-RS) in the downlink channel of the 5G NR commercial cellular networks. For this purpose, a highly accuracy time of arrival (ToA) estimation method is developed and a software defined radio (SDR) receiver is implemented. Furthermore, with the indoor positioning system  as shown in Fig.~\ref{fig:methodframe}, a lot of indoor field tests are carried out to verify the accuracy of the ToA estimation with real 5G NR signals.

In essence, the DM-RS is a broadcast signal transmitted by eNodeBS, which can be received by any 5G NR terminals for synchronization. Thus, the proposed positioning method emphasises the following advantages: Firstly, the number of users for positioning is not limited, because it is broadcast signals that is used for positioning. Hence, no extra bandwidth is required for positioning, when there are more users. This is contrast to the newly defined reference signals for positioning, such as the PRS/uplink CSI signals. Such a property is utmost important in the applications, such as emergent rescue, when the request for positioning has to be acknowledged by a big number of users simultaneously. In addition, this property can also help to provide the localization enhanced services in Software Defined Vehicular Networks~\cite{cite09-VN1},~\cite{cite09-VN2}. Secondly, the users can receive the signals without revealing any related information of its own. Thus, comparison with the network based positioning, our proposed method can preserve users' privacy. Considering the application in the field of IoT, our method is expected to be well suitable for the incentive mechanism design of crowdsensing for privacy-preserving~\cite{cite09-CS1},~\cite{cite09-CS2}. Thirdly, the signals are transmitted periodically, which is beneficial for users to continuously update its estimation of ToA and hence improve it localization accuracy. Finally, as a method implemented at the user side, it is relatively easy for users to fuse the positioning information with the information obtained from the other sources embedded in smartphones, such as, inertial sensors, sound, etc.

This paper's contributions are summarized as follows:
\begin{enumerate}[]
	\item A carrier phase based ToA estimation algorithm for the 5G NR signals has been proposed. The performance of ToA estimation has been thoretically analyzed.
	\item A solution of 5G NR software defined radio (SDR) receiver for the ToA estimation has been proposed, which includes coarse synchronization, multipath acquisition, delay tracking and carrier phase based ranging estimation.
	\item 
	The effectiveness of the proposed method has been verified by the indoor field tests in an office environment, where a commercial 5G NR base station (known as gNB) is installed. To the best of our knowledge, this is the first study of the indoor positioning with the commercial 5G NR signals in sub-6 GHz band.
\end{enumerate}

The rest of this paper is organized as follows: Section~\ref{sec:model} briefly overviews the 5G NR standard in the perspective of wireless positioning and presents the signal model with channel impairments. Section~\ref{sec:toa_method} describes the ToA estimation method designed based on carrier phase measurement. Section~\ref{sec:theory} derives the autocorrelation function used in our method and analyses the ToA tracking errors. The test bench used for sampling the commercial 5G NR signals and the testing scenario for indoor experiments are described in Section~\ref{sec:testbench}, where ranging results are demonstrated and discussed. Finally, in Section~\ref{sec:conclusion}, conclusions are summarized.

\begin{table}[!t]
	\renewcommand\arraystretch{1.8}
	\begin{center}
		\caption{Expandable OFDM parameter set of 5G NR}
		\label{table:Expandable_OFDM_parameter}
		\resizebox{83mm}{14mm}{
			\begin{tabular}{|c|c|c|c|c|}
				\hline
				\begin{minipage}{3cm} \vspace{1mm} \centering OFDM Parameter Set \vspace{1mm} \end{minipage} & 15 KHz & 30 KHz & 60 KHz & 120 KHz\\
				\hline
				\begin{minipage}{3cm} \vspace{1mm} \centering OFDM Symbol \vspace{1mm} \end{minipage} & 66.7 $\mu$s & 33.33 $\mu$s & 16.67 $\mu$s & 8.33 $\mu$s\\
				\hline
				\begin{minipage}{3cm} \vspace{1mm} \centering CP Length \vspace{1mm} \end{minipage} & 4.69 $\mu$s & 2.34 $\mu$s & 1.17 $\mu$s & 0.59 $\mu$s\\
				\hline
				\begin{minipage}{3cm} \vspace{1mm} \centering OFDM With CP \vspace{1mm} \end{minipage} & 71.35 $\mu$s & 35.68 $\mu$s & 17.84 $\mu$s & 8.91 $\mu$s\\
				\hline
				\begin{minipage}{3cm} \vspace{2mm} \centering \multirow{2}{*}{Frequency Band} \vspace{2mm} \end{minipage} &
				\multirow{2}{*}{0.45 $\sim$ 6 GHz} & \multirow{2}{*}{0.45 $\sim$ 6 GHz} & 0.45 $\sim$ 6 GHz & \multirow{2}{*}{24 $\sim$ 52.6 GHz}\\
				& & & 24 $\sim$ 52.6 GHz & \\
				\hline
				\begin{minipage}{3cm} \vspace{1mm} \centering Maximum Bandwidth \vspace{1mm} \end{minipage} & 50 MHz & 100 MHz & 200 MHz & 400 MHz\\
				\hline
			\end{tabular}
		}
	\end{center}
\end{table}

\section{5G NR standard and the signal model}~\label{sec:model} \subsection{Overview of 5G NR signals from the Perspective of Wireless Positioning}
\subsubsection{Physical Signal and the Time-Frequency Resources}
The 5G NR standard uses the orthogonal frequency division multiplexing (OFDM) modulation to achieve robust transmission in multipath scenarios. In 3GPP Release 15 of 5G NR standard, OFDM signals are specified by three parameters, namely, the number of subcarriers or the fast Fourier transform (FFT) size, the sampling period and the length of cyclic prefix (CP). In order to enable diverse services on a wide range of frequencies and deployments, 5G NR has a scalable OFDM numerology and  an extensible set of CP-OFDM parameters, as shown in Table~\ref{table:Expandable_OFDM_parameter}. 

\begin{figure}[t]
	\centering
	\includegraphics [width=\linewidth]{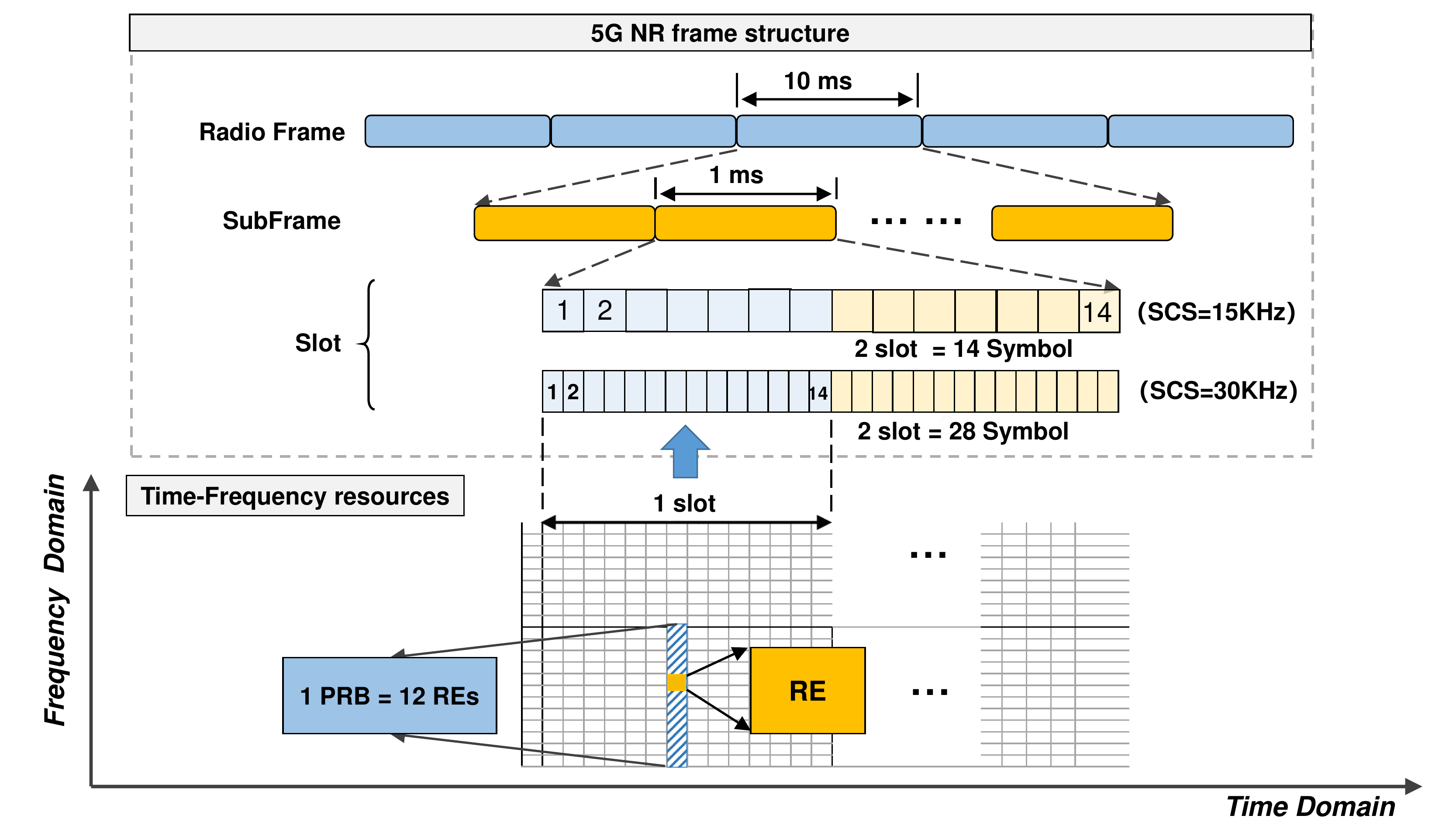}\\
	\caption{5G NR time-frequency resources and frame structure.}
	\label{fig:TFstructure}
\end{figure}

In 5G NR, physical layer uses time-frequency resources for transmission. The smallest physical time-frequency resource consists of one subcarrier in one OFDM symbol, which is defined as a resource element (RE). The transmissions are scheduled in group(s) of 12 subcarriers, defined as physical resource blocks (PRBs). The time-frequency resources is illustrated in Fig.~\ref{fig:TFstructure}, where the physical signal's transmissions are organized into signal frames, subframes, and slots in the time domain. Each signal frame has a duration of 10 ms and consists of 10 subframes with a subframe duration of 1 ms. A subframe is formed by one or multiple adjacent slots.
In this study, the parameter of 30~KHz subcarrier spacing was used in our experiments and thus, one subframe has 2 slots, and each slot has 14 OFDM symbols.
\begin{table}[t]
	\renewcommand\arraystretch{1.4}
	\begin{center}
		\caption{Resources within an SSB block for PSS, SSS, PBCH, and DM-RS for PBCH}
		\label{table:ResourcesSSB}
		\resizebox{90mm}{22mm}{
			\begin{tabular}{|c|c|c|}
				\hline
				\multirow{2}{*}{Channel or signal} & OFDM symbol number $l$ relative & Subcarrier number $k$ relative \\
				&  to the start of an SSB block & to the start of an SSB block \\
				\hline
				PSS & 0 & 56,57, ... ,182\\
				\hline
				SSS & 2 & 56,57, ... ,182\\
				\hline
				\multirow{2}{*}{Set to 0} & 0 & 0,1, ... 55,183,184, ... 239\\
				\cline{2-3}
				& 2 & 48,49, ... 55,183,184, ... 191\\
				\hline
				\multirow{2}{*}{PBCH} & 1, 3 & 0,1, ... ,239\\
				\cline{2-3}
				& 2 & 0,1, ... ,47,192,193, ... ,239\\
				\hline
				\multirow{3}{*}{DM-RS for PBCH} & 1, 3 & [0,4,8, ... 236] + V\\
				\cline{2-3}
				& \multirow{2}{*}{2} & [0,4,8, …44] + V\\
				& & [192,196, …236] + V\\
				\hline
			\end{tabular}
		}
	\end{center}
\end{table}
\begin{figure}[t]
	\centering
	\includegraphics [width=.9\linewidth]{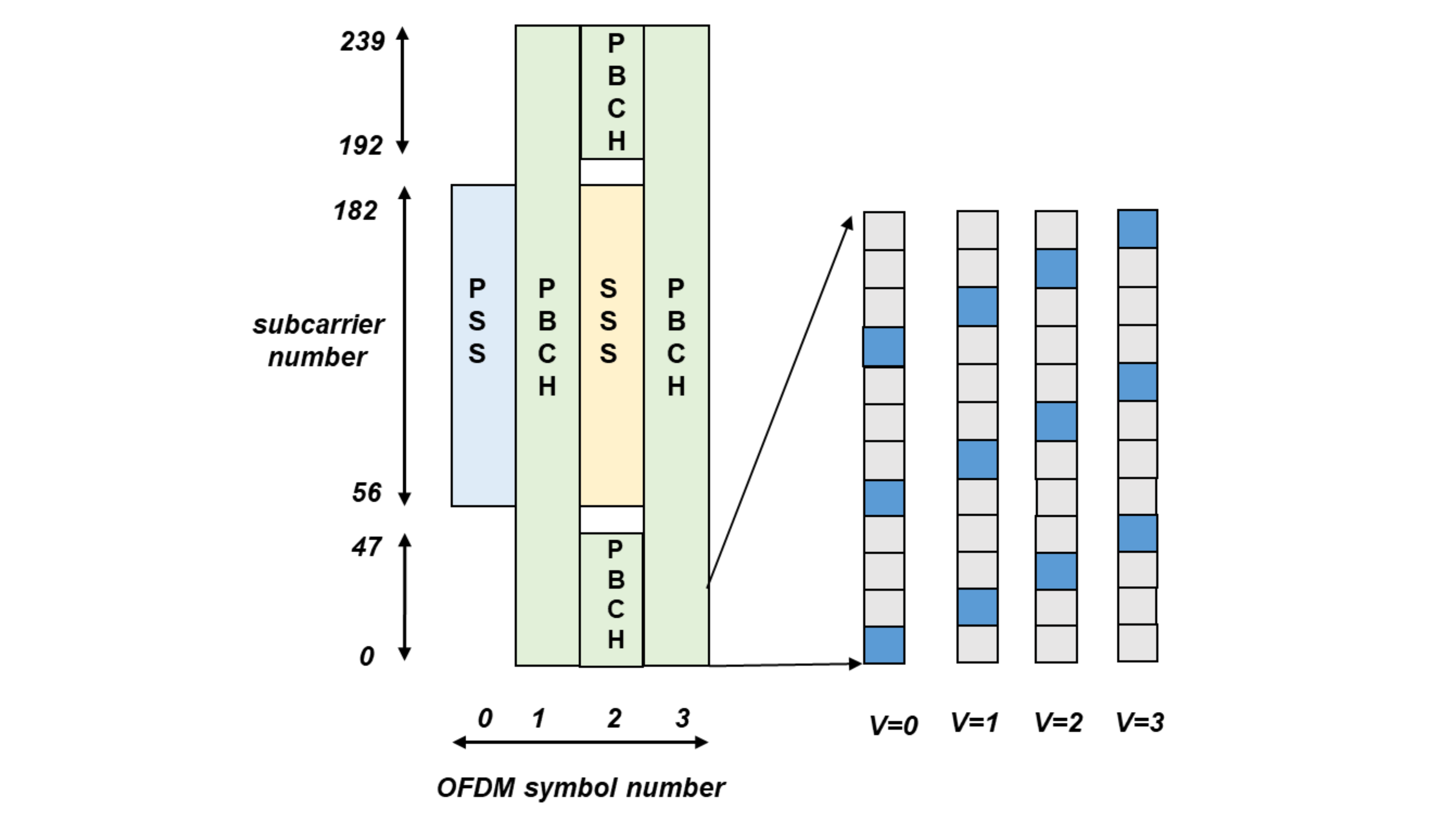}\\
	\caption{5G NR SSB structure.}
	\label{fig:SSB_structure}
\end{figure}
\subsubsection{Synchronization Signals in Downlink Physical Channels} 
There are typically four physical channels in the 5G NR downlink, which include the synchronization signal burst set (SS burst set), physical downlink shared channel (PDSCH), physical downlink control channel (PDCCH) and the control resource set (CORESET). 
The SS burst set has multiple synchronization signal blocks (SSB), which is transmitted periodically in a fixed time interval. The SSB mainly includes the primary synchronization signal (PSS), the secondary synchronization signal (SSS), the demodulation reference signal (DM-RS) and the physical broadcast channel (PBCH). In section TS 38.211 of 5G NR protocols~\cite{cite24}, PSS and SSS are generated by the pure BPSK-modulated $m$ sequence of length 127, and DM-RS is generated by the pseudo-random sequence, namely the gold-sequence of length 31. The protocols stipulate that there are 3 different PSS sequences, and each PSS sequence corresponds to 336 different SSS sequences. 
In the time domain, each SSB consists of 4 OFDM symbols, numbered in ascending order from 0 to 3, where PSS, SSS, DM-RS and PBCH are mapped to the symbols in SSB. Table~\ref {table:ResourcesSSB} shows the resources in SSB, where PSS and SSS correspond to the first and third symbols in the SSB, respectively. In the frequency domain, SSB occupies the synchronization signal burst set with 240 subcarriers, and PSS and SSS are located in the subcarrier numbers of 56 to 182. The time-frequency structure of the SSB block is shown in Fig.~\ref{fig:SSB_structure}, the blue REs are DM-RS scattered in PBCH. According to Fig.~\ref{fig:SSB_structure}, there are three types of DM-RS on each PRB of the PBCH, and the DM-RS has 4 frequency domain offsets related to the cell identities (CID).

From the perspective of wireless positioning, there are several properties of the SSB that can be utilized. Firstly, PSS and SSS in SSB are designed for synchronization, while DM-RS can be used for channel estimation. Thus, by exploiting the autocorrelation property of the PSS and SSS, and also, DR-MS with the boosted power, it is possible to achieve the accurate estimation for the ToA of the incoming 5G downlink signals. Secondly, the signals embedded in SSB are periodically transmitted in a fixed  duration, which enable the receiver to continuously track the arrival of signals, and thus improve the accuracy of the timing-based estimation for wireless positioning.
\begin{figure*}[t]
	\centering
	\includegraphics [width=\linewidth]{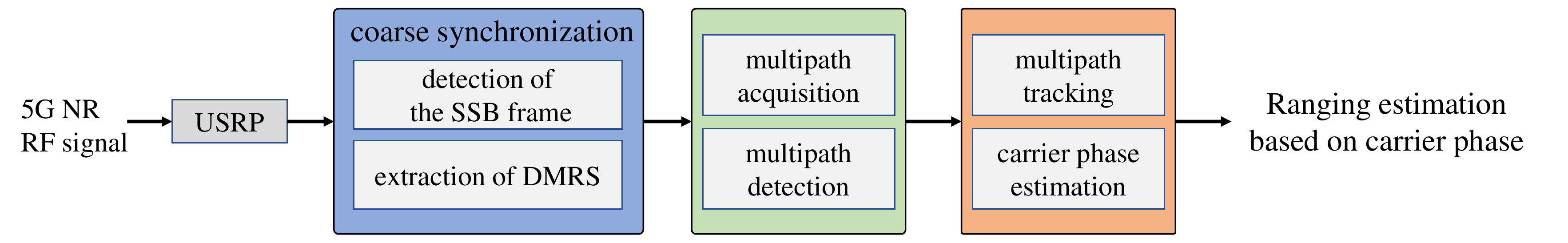}\\
	\caption{Block diagram of the SDR 5G NR DL receiver for the ranging estimation based on carrier phase estimation.}
	\label{fig:Block_diagram}
\end{figure*}
\subsection{Signal model}
Let us assume a 5G NR system consisting of $N$ subcarriers, among which $N_u$ subcarriers at the central spectrum are used for transmission and the other subcarriers at both edges form the guard bands.
Let~$\{c_n| n = 0,\cdots,N_u-1\}$ denote the modulated data or subcarrier symbol, where $n$ represents the subcarrier number. After the inverse fast Fourier transform (iFFT) operation and by adding the guard interval for every OFDM symbol, the samples of the transmitted baseband signal can be expressed as
\eqnum[eq_bb_tr]{
	s(k) =\frac{1}{\sqrt{N}}\sum_{n=0}^{N_u-1}c_n e^{j2\pi k n/N} \hspace{2ex} -N_g \leq k \leq (N-1)
}
where $N_g$ is the number of guard samples, and $j = \sqrt{-1}$.

In a typical indoor environment, wireless signals are commonly reflected or scattered by obstacles, which leads to multipath effect. Hence, we assume that the 5G NR signals experience frequency selective fading, and the channel impulse response (CIR) is of length $L$ with the complex channel gains of  $\{h_l\}$ and the corresponding path delays of $\{\tau_l\}$, for $l = 0, 1, \cdots,L-1$.

At the receiver, the received signals usually exist symbol-timing offset (STO), carrier-frequency offset (CFO), and sampling clock offset (SCO). To be more specific, STO is the time difference between the supposed and the real beginning of one OFDM symbol. CFO is caused by the mismatch between the frequency generated by the receiver's local oscillator and the carrier of the received signal. Finally, SCO is due to the mismatch of the sampling clock frequencies caused by the Doppler effect. After tracking all these factors into account, the received samples can be written as
\eqnum[eq:rec_sig_error]{
	r(k) = e^{j(2\pi k \Delta f/N + \varphi)}\sum_{l=0}^{L-1} h_l s(k -\tau_l) + n(k)
}
where $n(k)$ is the sample of zero-mean complex Gaussian noise
distributed with zero mean and variance of $\sigma^2$, $\Delta f$ is the CFO normalized by the subcarrier spacing and $\varphi$ is an arbitrary carrier phase. The timing point at the start of the FFT window is determined by the timing synchronization to be at the sample $r(\epsilon)$, where $\epsilon$ is a timing offset in the unit of OFDM samples.

It has been recognized that the channel effect and the timing/frequency errors in (\ref{eq:rec_sig_error}) impose important impact on the signals demodulation \cite{cite25}. For the purpose of communication, 5G NR receivers extract the timing measurements and recover the frequency offset from the received signals, which results in a synchronization problem. Although multicarrier communication systems have stringent requirements for the timing and frequency synchronization for achieving reliable communications, as what are required for positioning and navigation, it is still important to attain a finer synchronization to obtain a finer time delay estimation of the received signals, so that sufficient localization accuracy can be obtained. In the following section, we  described our proposed scheme for the ToA estimation of the 5G NR signals, which is critical for achieving high accuracy.

\section{ToA Estimation based on Carrier Phase Measurement for 5G NR positioning}~\label{sec:toa_method}
The objective of the ToA estimation of 5G NR signal is to find the starting point of the incoming OFDM symbols as accurately as possible. Since SSB is periodically transmitted in DL, in this work, the ToA estimation for 5G NR positioning is to find the starting time of the SS signals.
Considering the time and frequency errors as well as the multipath effect as mentioned with (2), four steps are operated to estimate the ToA of 5G NR signals, namely the coarse synchronization, the multipath acquisition of incoming signals, the fine estimation with delay tracking and the ToA measurement with carrier phase estimation. Fig.~\ref{fig:Block_diagram} shows the block diagram of the receiver. Let us below describe the four steps.
\subsection{Coarse Synchronization} \label{section_cync}
This step includes estimating the coarse start of the SSB frame, detecting the cell ID, and extracting the DM-RS symbols from the SSB, which are detailed as follows.
\subsubsection{Detection of the SSB Frame}
The start of the SSB frame is coarsely estimated by detecting the PSS and SSS. In this work, we introduce the correlation method to correlate the received samples with all the possible choices of PSS and SSS.
To detect the PSS, let the candidate sequences of PSS in time domain be expressed as $s_\text{PSS}(n) = \mathcal{F}^{-1}\{\text{S}_\text{PSS}(n)\}$
where $\mathcal{F}\{\cdot\}$ is the discrete-time transform operator. Then, the correlation outputs can be expressed as 
\eqnum[eq_dect]{
	\Phi_\text{PSS}^{m_2}(k) & = \sum_{n=0}^{N_u-1} s^{m_2}_\text{PSS}(n)\cdot r^*(n+k), m_2\in \{ 0,1,2\}
}
where $n$ is the correlation lag, $\{r(n)\}$ are the received samples and $N_u$ is the symbol length of PSS. 
Based on the three possible sequences of PSS, the one giving the maximum value yields the estimated sequence for the PSS. Correspondingly, the coarse start of the SSB frame can be estimated. Similarly, the same correlation method is applied to detect the SSS.

\subsubsection{Extraction of DM-RS}
Let the estimated sequence number of the current SSS symbol be denoted as $\hat{m}_1 $, where~$\hat{m}_1\in\{0,1,...,335\}$, while the estimated PSS sequence number be expressed as $\hat{m}_2$. Then, the CID can be computed as
\eqnum[eq_pci]{
	N_{ID}^{cell}=3\times \hat{m}_1 +\hat{m}_2
}

With the aid of the coarse timing estimation, the frequent domain symbols in SSB can be derived by first removing the cyclic prefix and then applying the FFT on the samples belonging to a same OFDM symbol. Then, the DM-RS can be extracted from the subcarriers with the starting subcarrier index decided by the CID estimated by (\ref{eq_pci}).

\subsection{Multipath Acquisition}\label{sec:multipath_acquisition}
When the DM-RS is obtained, the CSI is able to be obtained, which can be explained by the receiver to demodulate data message. However, to achieve the positioning and navigation of sufficient accuracy, it is necessary to have more accurate symbol timing estimation. In this work, the acquisition and tracking steps are applied to achieve this objective.

The acquisition step aims to reduce the residual errors
resulted from the coarse synchronization and provides a more accurate start point as the initial value for the tracking step. 
In indoor wireless communication scenarios, signals experience the multipath fading, which is sometimes very severe. To improve the timing accuracy, we try to detect the individual multipaths and estimate their delays.

Denote $ c_{i}(p_s)$ as the local replica of the DM-RS in the $i$th SSB, where $p_s$ are the indices of the subcarries of all the DM-RS in $\mathcal{P}_s$. Let $d_i(p_s)$ denote the received symbols of the DM-RS subcarriers and $N_p$ be the total number the DM-RS. According to the time shift property of the Fourier transform, i.e. 
\eqnum[fft-delay]{
	\mathcal{F}\{r(k \pm \tau)\} = d(n) \cdot \exp(\pm j \frac{2\pi n \tau}{N})
}
we can know that the delay estimation is equivalent to phase estimation in frequency domain. Therefore, the multipath acquisition problem is to find $\hat{h}_{i,l}$ and $\hat{\tau}_{i,l}$ that minimize the error between the received DM-RS and the local replica, described as
\eqnum[eq:mp]{
	\{\hat{h}_{i,l}, \hat{\tau}_{i,l}\} = \arg \min_{h_{i,l}, \tau_{i,l}} \sum_{p_s \in \mathcal{P}_s} \norm{d_i(p_s) - \sum_l h_{i,l} e^{j \frac{2\pi p_s \tau_l}{N}}c_i(p_s)}
}
where $l$ is the number of multipaths, $h_{i,l}$ is the channel coefficient of the $l$-th path in the time of the $i$-th SSB symbol and ${\tau}_{i,l}$ is the corresponding time delay.

To solve the problem of (\ref{eq:mp}), we introduce the matching pursuit (MP) method \cite{cite26}. The principle of MP is to utilize the sparseness of the CIR and aims to find the linear combination of the lowest dimensional for the delayed versions of the possible transmitted symbol sequences to represent the received signal~\cite{cite27}.
In order to simultaneously estimate the delays and complex channel coefficients, an order-recursive least square MP (LS-MP) algorithm has been verified to be effective in our previous work~\cite{cite28}. To apply the method for the current problem, let us first rewrite (\ref{eq:mp}) in matrix form as
\eqnum[eq:mp-2]{
	\{\hat{h}_{i,l}, \hat{\tau}_{i,l}\} = \arg \min_{h_{i,l}, \tau_{i,l}} \norm{\vct{d_i}  - \vct{C_i} \cdot \vct{h_i}}
}
where $\vct{C_i}$ is  a ($N_p \times L $) matrix, $\vct{d_i}$ is a ($N_p \times 1$) vector with the element of $d_i(p_s)$ and $\vct{h_i}$ is a ($L\times 1$) vector. Therefore, a solution $\hat{\vct{h_i}}$ may be viewed as the coefficient vector associated with the representation  of $\vct{d_i}$ in terms of the columns of $\vct{C_i}$.

In order to estimate the $L$ path delays, let us define the time delay sequence as $\Upsilon = [0, \Delta \tau, \cdots, (N_{\tau} -1 ) \Delta \tau] $, where $\Delta \tau$ is the time interval for the delay estimation and  $N_\tau$ is the total number of possible delays. Then, a matrix $\vct{C_i}^\prime$ is constructed with the row element expressed as $\vct{c_i(p_s)} = c_i(p_s) \cdot \exp\{j\frac{2\pi p_s \Upsilon}{N}\}$. The objective of the order-recursive LS-MP algorithm is to find iteratively the columns of $\vct{C_i}^\prime$ that best matches the CIR estimates. After an iteration, the CIR vector is corrected by removing the pulse of the retained path and a new delay is estimated from $\Upsilon$.
The iterative search process continues until a given number, say $L$, is extracted, or the residual power of the signal is smaller than a  pre-set threshold $\Gamma_{\text{acq}}$. 
\subsection{Multipath Tracking}
Once the multipaths are acquired, with the accuracy on the order of $\Delta \tau$, multiple tracking loops are then implemented to filter the time delay so as to achieve even more accurate ToA estimation of the multipaths. Similar to the multipath acquisition, multipath tracking is also implemented in the frequency domain based on the time shift property of Fourier transform. Multiple Delay-locked loops (DLLs) for multipath tracking are implemented as follows.

After the multipath acquisition, the time delays of the $L$ multipaths in the $i^\text{th}$ SSB are expressed as $\{\hat{\tau}_{i,0},\cdots,\hat{\tau}_{i,L-1}\}$, with the corresponding channel coefficients expressed as $\{\hat{h}_{i,0},\cdots,\hat{h}_{i,L-1}\}$. To track a specific path, the receiver's timing is first adjusted according to the estimated normalized symbol timing error, which can be expressed as 
\eqnum[eq:dll1]{
	\hat{c}_{i,l} (p)= e^{-j\frac{2 \pi p \hat{\tau}_{i,l} }{N}}c_{i}(p),  l = 0, 1, \cdots,L-1
}
where $\hat{\tau}_{i,l}$ is the normalized delay of the $l^\text{th}$ path obtained by the multipath acquisition step. Then, the received DM-RS are cross-correlated with the locally generated early and late reference DMRS, yielding
\eqnum[eq:dll3]{
	R_{i,l} (\epsilon) &= \frac{1}{N_p}\sum_{p \in \mathcal{P}_{s}} {z}_{i,l}(p)\cdot \xi_{i,l}^*(p,\epsilon)\\
	R_{i,l} (-\epsilon) &= \frac{1}{N_p}\sum_{p \in \mathcal{P}_{s}} {z}_{i,l}(p)\cdot  \xi_{i,l}^*(p,-\epsilon)\\
}
where ${\xi}_{i,l}(p,\epsilon) = e^{+j\frac{2 \pi p \epsilon }{N}} \hat{c}_{i,l}(p)$ and $\xi_{i,l}(p,-\epsilon) = e^{+j\frac{2 \pi p (-\epsilon)}{N}} \hat{c}_{i,l}(p)$, and $ \epsilon~(0 < \epsilon < 1/2)$ is the advanced (and retarded) interval, which is normalized by the OFDM sample interval, while $\hat{z}_{i,l}(p)$ is given by
\eqnum[eq:hat_z]{
	z_{i,l}(p) &= d_i(p) - \Sigma_{j=0 }^{l-1}\hat{h}_{i,j} \cdot  e^{-j\frac{2 \pi p \hat{\tau}_{i,j} }{N}}\cdot c_i(p)\\
}

Assume the Early-Minus-Late Power (EMLP) discriminator for DLL tracking. The normalized discriminator is expressed as
\eqnum[dll4]{
	a_{i,l} (\epsilon) =  \frac{1}{k_{i,l}}\left(\abs {R_{i,l} (\epsilon)}^2 - \abs {R_{i,l}(-\epsilon)}^2\right)
}
where the normalization factor $k_{i,l}$  is used to keep $a_{i,l}(\epsilon) \approx \epsilon_{i,l} $, when $\epsilon_{i,l} \rightarrow 0$. By smoothing the discriminator $a_{i,l}(\epsilon)$ with a loop filter, the delay estimate to the $(i+1)^\text{th}$ SSB symbol is finally updated to $\hat{\tau}_{i+1,l} = \hat{\tau}_{i,l} + \tilde{\epsilon}_{i,l}$,
where $\tilde{\epsilon}_{i,l}$ is the output of the loop filter.

Based on the tracked delay $\hat{\tau}_{i+1,l} $,  $\hat{c}_{i+1,l} (p)$ can be updated according to (\ref{eq:dll1}). Let the updated vector be expressed as $\vct{c}_{i+1,l} = [\hat{c}_{i+1,l} (0), \cdots,\hat{c}_{i+1,l} (M)]^T$, and let $\vct{z}_{i,l} = [\hat{z}_{i,l} (0), \cdots,\hat{z}_{i,l} (M)]^T$. Then, the corresponding channel coefficients can be updated to
\eqnum[mp_a]{
	\hat{h}_{i+1,l} = (\vct{c}_{i+1,l}^T \cdot \vct{c}_{i+1,l}) \cdot \vct{c}_{i+1,l}^T\cdot \vct{z}_{i,l}
}

The multipath tracking process executes the above described DLL iteratively for each of the multipaths. The iteration for a specific multipath stops, if the remaining power is small or than a pre-set threshold. 
\subsection{Carrier Phase Based Range Estimation}\label{sec:carrier_phase_estimation}
Finally, the carrier phase can be simply derived from the first arrived path with the coefficient $h_{i,0}$, i.e. $\varphi_{i} = \angle h_{i,0}$. Correspondingly, the ranging estimation for a pedestrian of indoor navigation is calculated as
\eqnum[dll9]{
	\delta_i = (\frac{\varphi_{i}-\varphi_{i-1}}{2\pi}) \lambda 
}
where $\lambda$ is the wavelength of the carrier, which for the 5G NR is about 0.12 mm. According to~\cite{cite30}, the speed of a pedestrian walking indoor is usually less than 3 m/s. Since the update period of 5G NR is 20 ms, the phase difference between two adjacent SSB does not exceed one wavelength. Therefore, there is no ambiguity problem existing in the indoor pedestrian navigation.
\subsection{Algorithm Description}
In summary, the carrier phase based ToA estimation algorithm for 5G NR positioning  can be stated as follows.
\begin{algorithm}[t]
	\caption{Carrier Phase based ToA Estimation for 5G NR Positioning}\label{AlgorithmCPR}
	\textbf{1. Coarse synchronization of 5G NR}
	\begin{itemize}
		\item  Coarse timing estimation of the SSB frame by detecting both PSS and SSS according to~(\ref{eq_dect});
		\item  Extraction of DM-RS according to~(\ref{eq_pci}).
	\end{itemize}
	\textbf{2. Multipath acquisition} 
	\begin{itemize}
		\item Applying the MP algorithm to acquire the time delay of the $l$-th path in the  $i$-th SSB symbol, i.e.,  $\hat{\tau}_{i,l}$, according to~(\ref{eq:mp});
		\item Iteratively acquiring the time delays of the multipaths until a given number $L$ of multipaths are identified, or when the  power of the residual signal is smaller than a pre-set threshold. 
	\end{itemize}
	\textbf{3. Multipath tracking}
	\begin{itemize} 
		\item Applying the DLL in frequency domain to track the arrival time of the $l$-th path, updating the delay estimation to $\hat{\tau}_{i+1,l}$ according to~(\ref{eq:dll1}) - (\ref{dll4}), and updating on the corresponding channel coefficient $\hat{h}_{i+1,l}$ according to~(\ref{mp_a});
		\item Iteratively tracking the time delays of the individual multipaths until the residual power of the received signal is smaller than a threshold. 
	\end{itemize}
	\textbf{4.Carrier phase ranging estimation}
	\begin{itemize}
		\item Calculating the carrier phase of the first arrived path as $\varphi_{i} = \angle h_{i,0}$;
		\item Ranging estimation based on the carrier phase according to~(\ref{dll9}).
	\end{itemize}
\end{algorithm}

\section{Performance Analysis OF ToA estimation}~\label{sec:theory}
\subsection{Ideal Autocorrelation Function (ACF) of DM-RS in 5G NR PBCH}
Without considering noise, the normalized ideal ACF of DM-RS sequence is given by
\eqnum[eq:corr]{
	R_i^{o}({\epsilon}) &= \frac{1}{N_p}\sum_{p \in \mathcal{P}_{s}} {c}_i(p) \cdot \left(c_i(p) \cdot e^{-j\frac{2 \pi p \epsilon }{N}}\right)^*\\
	& = \frac{1}{N_p}\sum_{p \in \mathcal{P}_{s}} {c}_i(p) \cdot c_i^*(p) \cdot e^{j\frac{2 \pi p \epsilon }{N}}
}
In 5G NR, the pilots are inserted every $\kappa$ subcarriers. Denote $\E\left[{c}_i(p) \cdot c_i^*(p)\right] = A$, where $\E[\cdot]$ denotes statistical expectation. Then, (\ref{eq:corr}) can be simplified to
\eqnum[eq:corr2]{
	R_i^{o}(\epsilon) = \frac{1}{N_{p}} A e^{j\pi\epsilon\frac{\left[2p_i(0) + \kappa (N_{p}-1)\right]}{N}} \frac{\sin \pi \kappa N_{p} \epsilon /N}{\sin \pi \kappa \epsilon /N}
}
where $p_i(0)$ is the start index of the DM-RS in 1st or 3rd symbol of the $i$-th SSB. Furthermore, since $N>> 1$  (in the case of the DM-RS in 5G NR, $N = 256$)  and the values of $\epsilon$ is small within several samples, the following approximation of $\sin \frac{\pi \kappa \epsilon}{N} \approx \frac{\pi \kappa \epsilon}{N} \rightarrow 0$ is valid. Therefore, the correlation function (\ref{eq:corr2}) can be approximated as
\eqnum[eq:corr3]{
	R_i^{o}(\epsilon) = A e^{j\pi \alpha \epsilon} \textrm{sinc}(\pi \beta \epsilon )
}
where $\beta = \kappa \frac{N_{p}}{N}$, $\alpha = [2p(0) + \kappa (N_{p} -1)]/N$. 
Specifically, for the DM-RS of 5G NR, we have $\kappa = 4, p_0  \in \{1,2,3,4\}, Np = 60$, which give $\alpha \approx  1$ and $\beta = 0.9375$.

\begin{figure}[t]
		\centering
		\includegraphics[width=.85\linewidth]{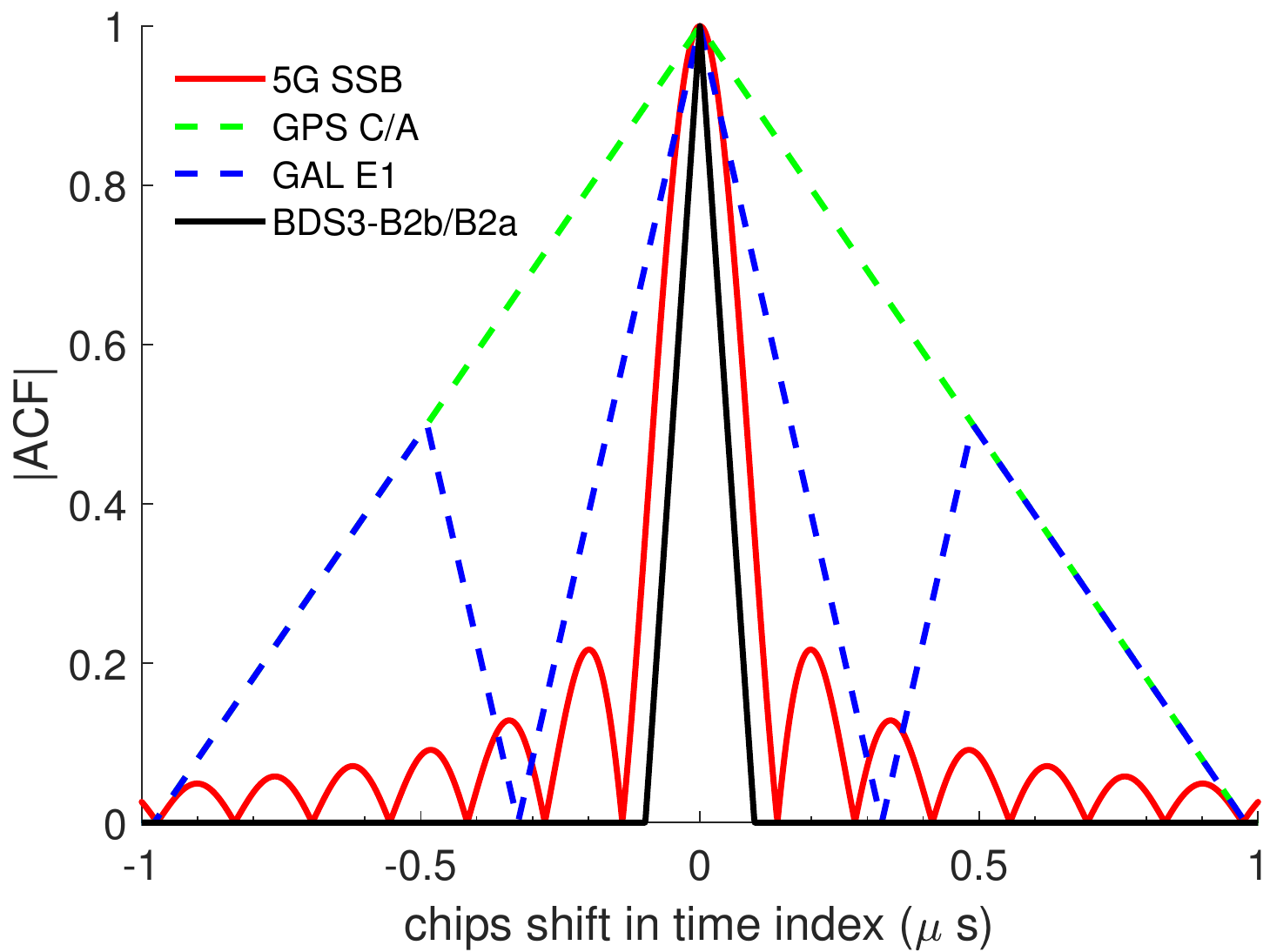}
		\caption{Ideal ACF of 5G SSB signal}
		\label{fig:IdealACF}
\end{figure}

The ideal ACF of DM-RS is shown in Fig.~\ref{fig:IdealACF}. For comparison, the ACFs of GPS C/A code, the Galileo E1 and BDS3-B2b/B2a code are also presented. It is observed that, for all the considered signals, the mainlobe of the ACF of BDS3-B2b/B2a is the narrowest, as the result that it has the largest bandwidth. When comparing with  GPS L1 and C/A Galileo E1,  the mainlobe of the ACF of 5G NR SSB is narrower by about $\frac{1}{7}$, which suggests an increased possibility of detecting two multipaths, especially, when the arrival time is larger than one chip of the 5G NR SSB, which is about 0.13 $\mu$s.

\begin{figure}[t]
		\centering
		\includegraphics[width=.75\linewidth]{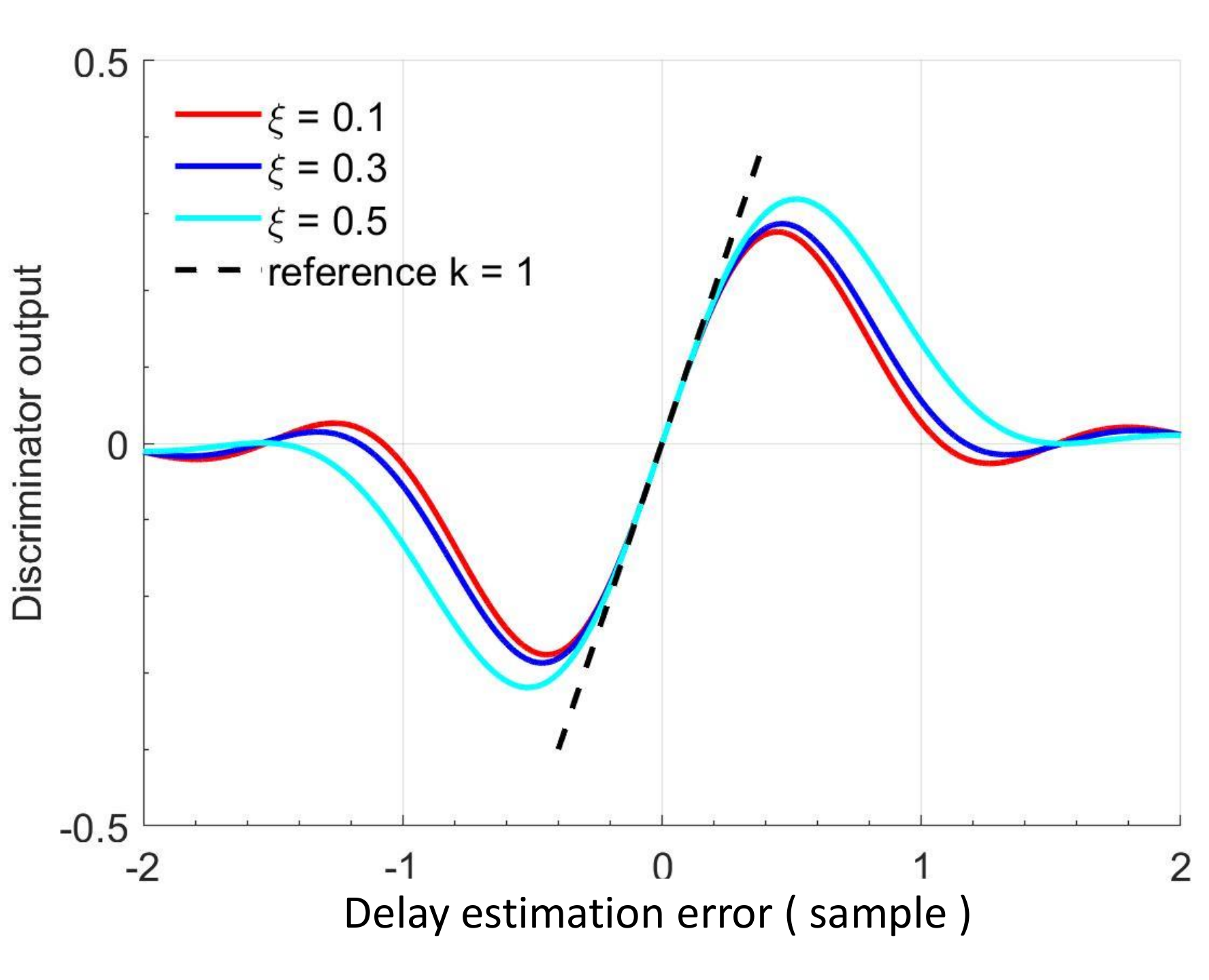}
		\caption{DLL S curve.}
		\label{fig:DLLScurve}
\end{figure}
\subsection{S Curve of the EMLP Discriminator}
Considering an additive white Gaussian noise (AWGN) channel with sufficient SNR,  S curve can be used to find the tracking errors of the EMLP discriminator. According to~\cite{cite29}, the S-curve $S(\epsilon,\xi)$ is expressed as
\eqnum[eq_a_useful]{
	S(\epsilon,\xi) 
	=  A^2\bigg(\textrm{sinc}^2(\pi \beta (\epsilon - \xi)) - \textrm{sinc}^2(\pi \beta (\epsilon + \xi))\bigg)
}
To get a normalized S-curve, the normalization factor $k_d$ can be derived as
\eqnum[eq_s_k]{
	k_d &= \frac{\partial S(\epsilon,\xi)}{\partial \epsilon} \vert_{\epsilon = 0}\\
	&= \frac{2A^2}{\pi^2\beta^2\xi^3}\bigg[\pi\beta\xi\sin(2\pi\beta\xi)  + \cos(2\pi\beta\xi) - 1  \bigg]
}
Fig.~\ref{fig:DLLScurve} illustrates the normalized S-curves $S_\text{norm}(\epsilon,\xi)$ with different settings of $\xi$.  Normally, in the DLL tracking, $\xi = 1/2$.

\section{Test bench for 5G NR signal sampling}~\label{sec:testbench}
A SDR test bench has been built for signal sampling and recording, which is based on USRP, a low-cost, flexible and open-source device originally developed by Ettus Research LLC~\cite{cite32}. Specifically, the USRP is responsible for the functions of clock generation, digital-analog signal interfacing, host processor interfacing, power management, up/down-conversion, analog filtering, and other analog signal processing. For the test bench, we used a USRP B210,  which has a main  board integrated with a Spartan 6 XC6SLX150 FPGA and an AD9361 RFIC conversion transceiver. It allows to receive the signals with the carrier frequency in the range between 70 MHz and 6 GHz, as well as a bandwidth of up to 56 MHz. The USRP B210 has amplifier with the adjustable gain from 0 to 31.5 dB, and a noise figure between 5 and 7 dB. Such configurations can be well adapted for the sampling of the 5G NR SSB signals.

In order to allow the local sampling clock to be synchronized with the GPS time, a GPS Disciplined Oscillator (GPSDO) module~\cite{cite33} is attached with the USRP B210, which provides a 10 MHz reference input and allows the master oscillator to be monitored by GPS. According to the specifications, the accuracy of the 1 PPS of the GPSDO is within $\pm{50}$ ns of the UTC RMS (1-$\sigma$), after the GPS is locked. 
\begin{figure}[t]
	\centering
	\begin{minipage}{0.5\textwidth}  
		\centerline{\includegraphics[width=\linewidth]{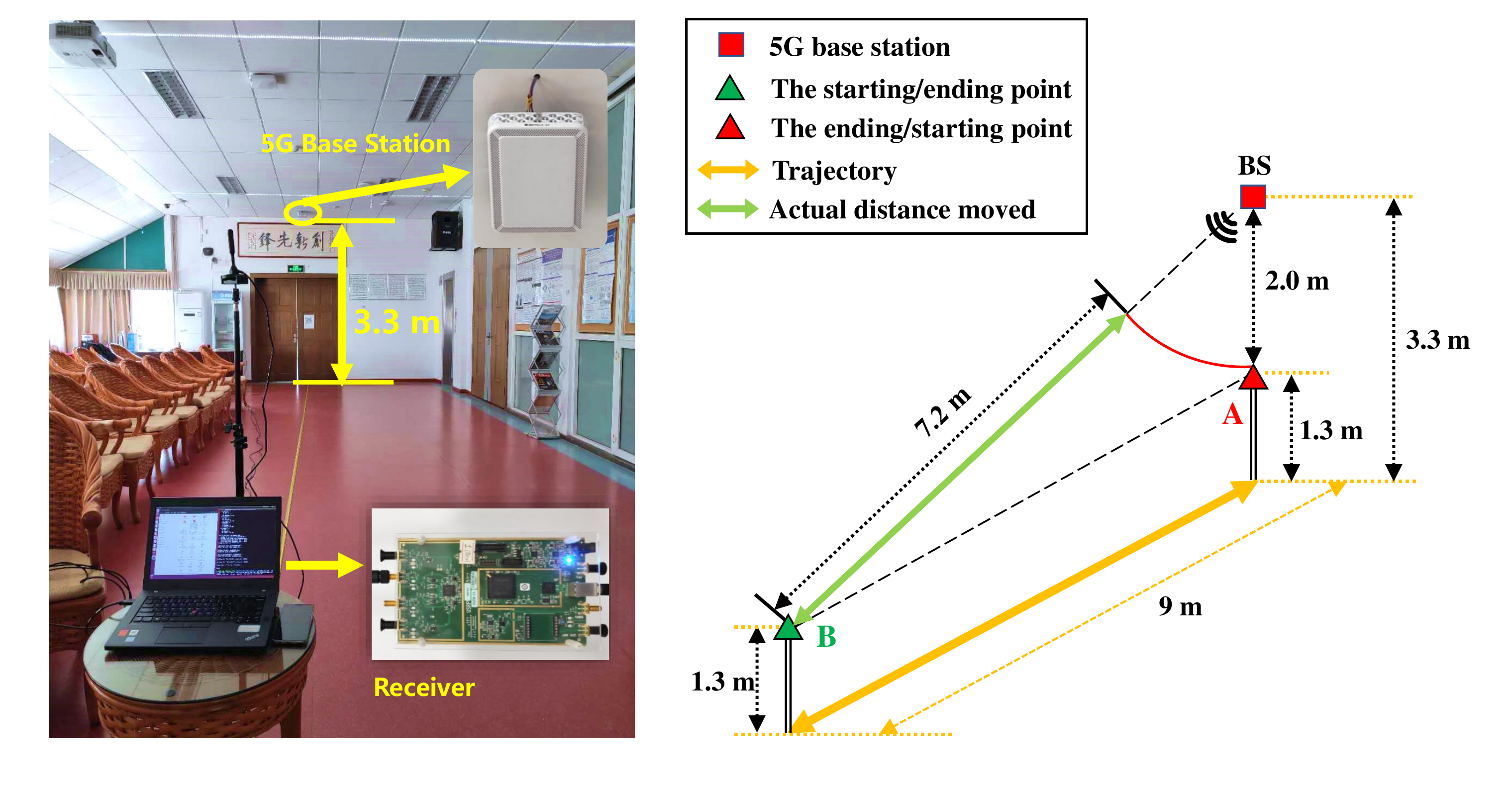}}
     	\centerline{(a)\space\space\space\space\space\space\space\space\space\space\space\space\space\space\space\space\space\space\space\space\space\space\space\space\space\space\space\space\space\space\space\space\space(b)}
    \end{minipage}
    \vfill
    \begin{minipage}{0.5\textwidth}  
	    \centerline{\includegraphics[width=.9\linewidth]{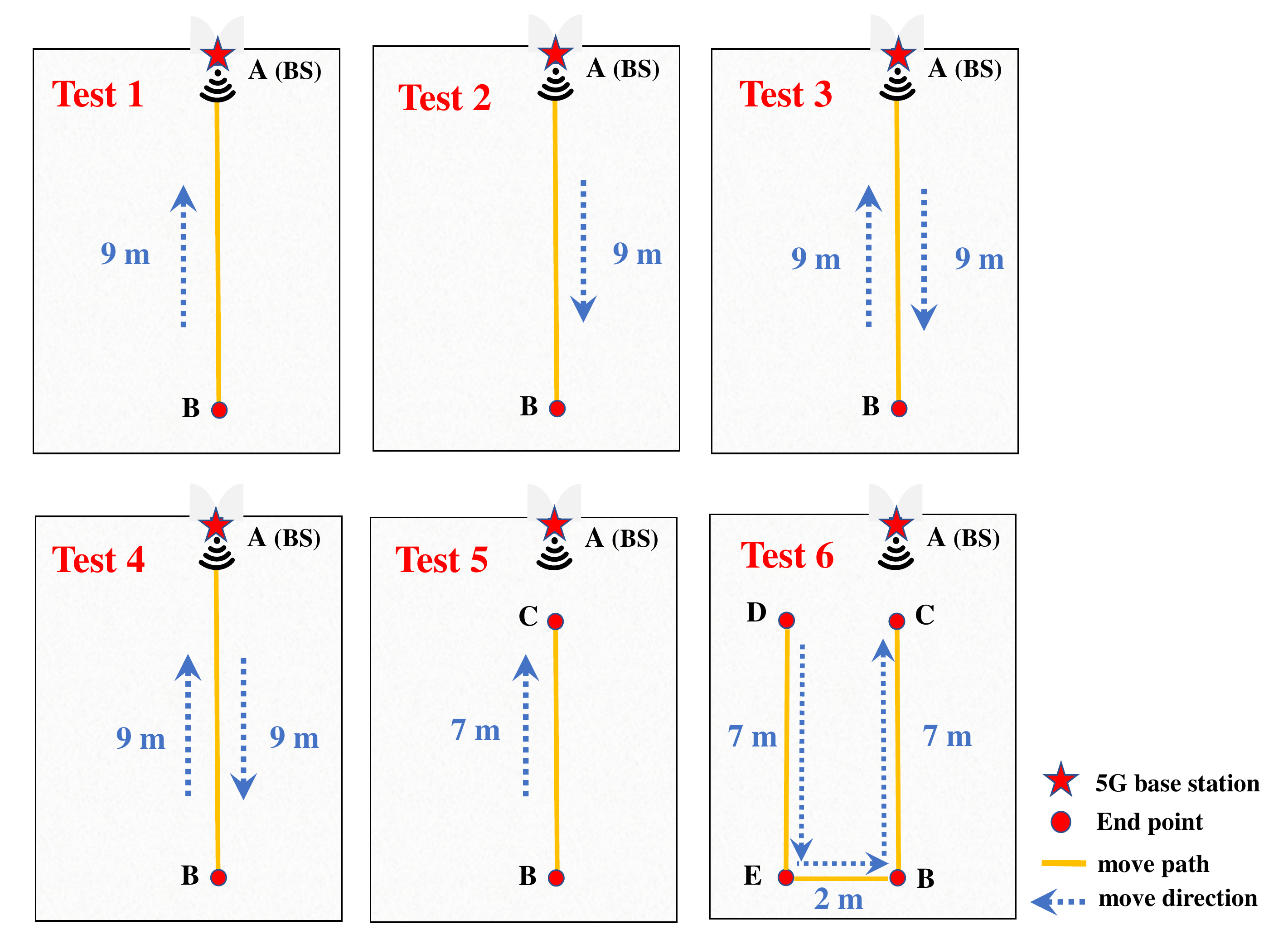}}
     	\centerline{(c)}
    \end{minipage}
    \caption{(a) USRP based SDR test bench. (b) 3D illustration of one indoor field test scenario. (c) 2D illustration of all the dynamic indoor field test scenarios.}
    \label{fig:scenario}
	\begin{minipage}[t]{.5\textwidth}
		\centering
		\includegraphics[width=.7\linewidth]{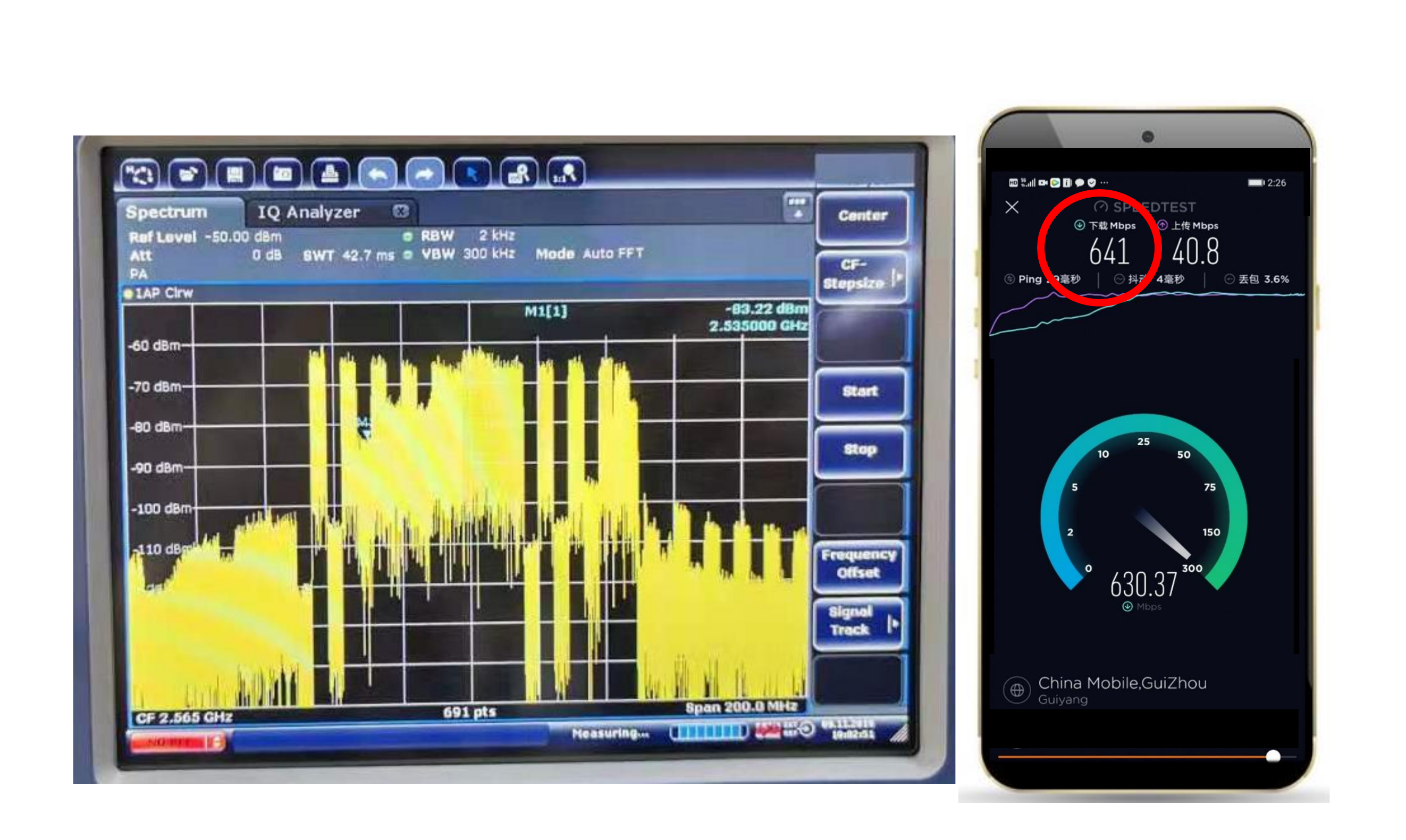}
		\caption{The snapshot of the signal spectrum and the realtime transmission speeds of 5G NR in the indoor field test.}
		\label{fig:testBench}
	\end{minipage}
\end{figure}
\section{Indoor Field Tests}~\label{sec:fieldtests}
In order to verify the developed algorithms and evaluate the performance of the carrier phase tracking of the real 5G commercial signals, indoor field tests were carried out in an office building at Wuhan University. In our indoor test environment, there is only one 5G NR gNB operated by China Mobile that can be heard. Therefore, we focus on evaluating the performance of the ToA estimation and of the carrier phase based ranging with 5G NR signals.
Fig.~\ref{fig:scenario} (a) shows the SDR test platform and the indoor field test scenario. Fig.~\ref{fig:testBench} shows one snapshot of the signal spectrum of the downlink transmission and the real-time transmission data rate in the indoor field test, where the downlink data rate is about 641 Mbps, while the uplink data rate is about 40.8 Mbps. Table~\ref{table:tests_parameters} describes in detail the relevant parameters of the gNB and hardware platform used in the field tests. In Table~\ref{table:tests_parameters}, the parameters of the gNB and of the transmitted signals are determined by the 5G NR protocol TS 38.104~\cite{cite33-2} and selected by a specific operator, the China Mobile in our case. According to the operator, the signal for commercial use is in Case C on band n41 and the corresponding subcarrier spacing is 30 KHz. At the receiver, we set the bandwidth to 10 MHz so as to receive most of the signal power of the SSB. During the stage of multipath acquisition, we set $\Delta \tau = 0.1$ and the threshold to $80\%$, so that most of the multipaths can be acquired and the first path can be detected with high reliability. In the DLL, the updating time is set to 20 ms, which is equal to the time interval of 5G NR SS/PBCH signal. Both static tests and dynamic tests are carried out. Specifically, during the static tests of ToA estimation, the bandwidth of DLL is set to 25 Hz to avoid too much smoothing on the ToA estimation and only observe the emitter clock behavior. By contrast, during the dynamic tests, the bandwidth is set to 0.5 Hz to ensure the loop convergence.

\begin{figure}[t]
	\centering
	\begin{minipage}[t]{.5\textwidth}
		\centering
		\includegraphics[width=.8\linewidth]{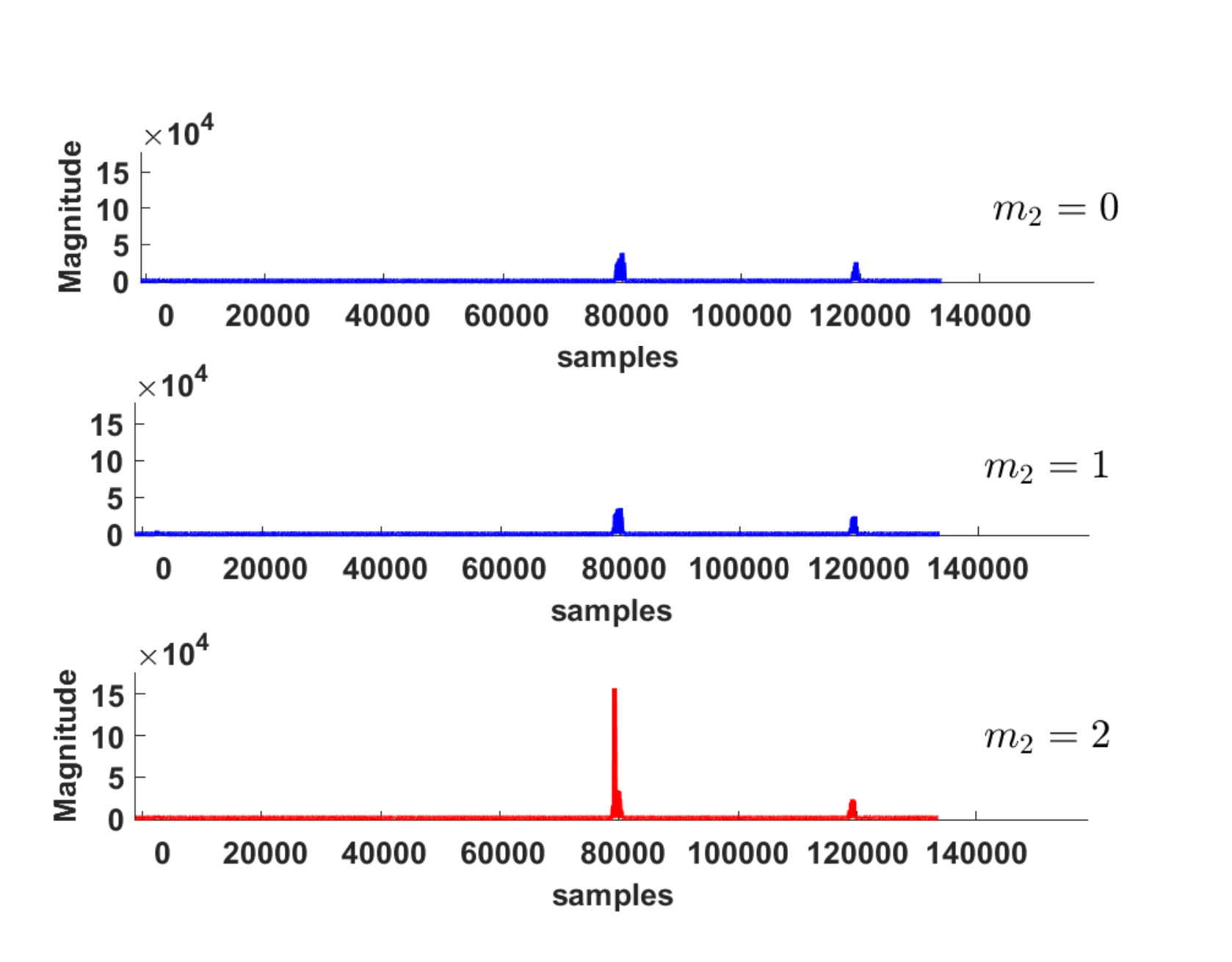}
		\caption{Coarse synchronization results based on the detection of PSS.}
		\label{fig:PSSDetect}
	\end{minipage}
	\begin{minipage}[t]{.5\textwidth}
		\centering
		\includegraphics[width=.75\linewidth]{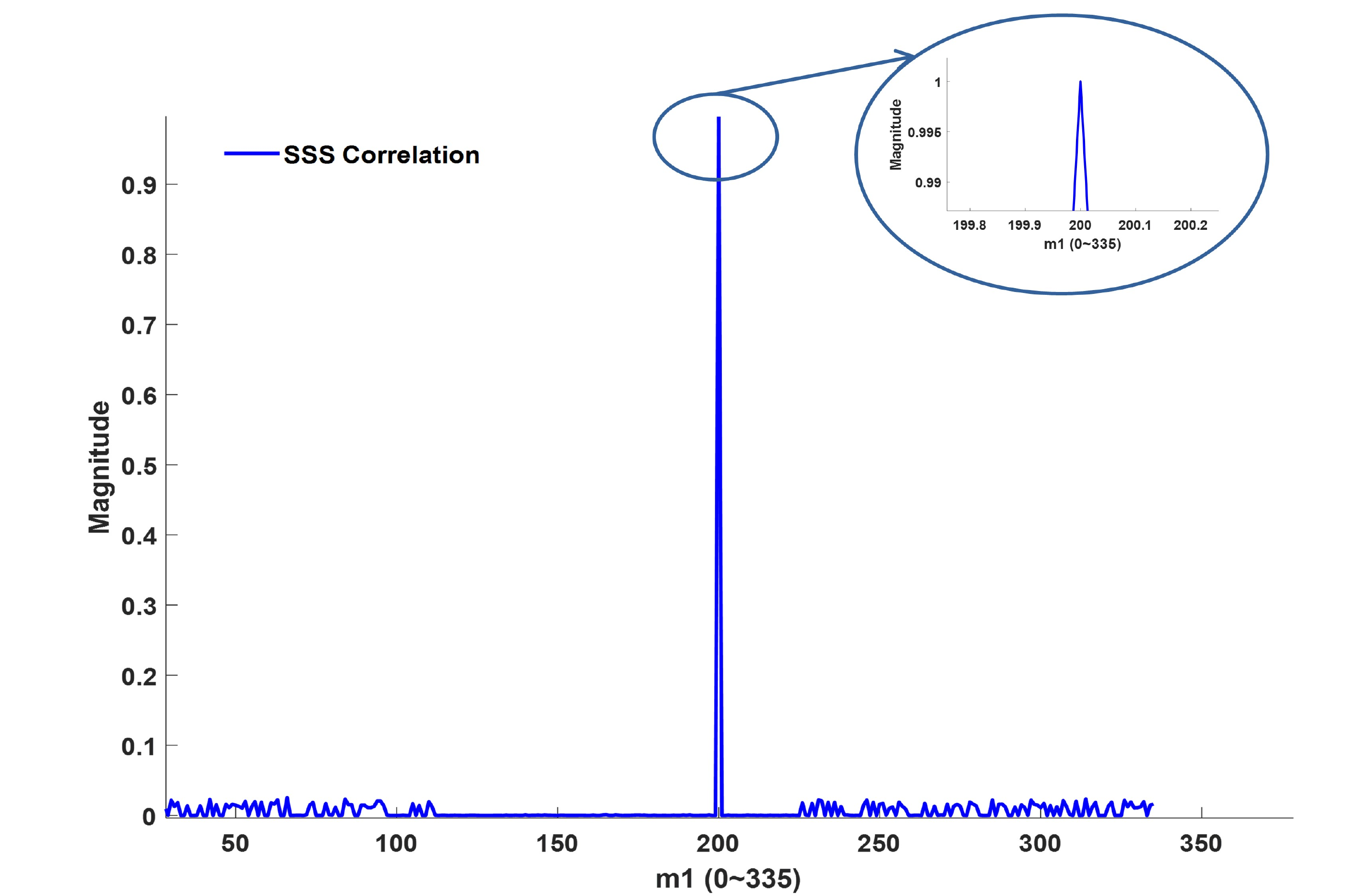}
		\caption{The detection of SSS.}
		\label{fig:SSSDetect}
	\end{minipage}
\end{figure}

\subsection{Static Tests}
For the static tests, the receiver is kept static on a desk located at point B of Fig.~\ref{fig:scenario} (a). The signals are sampled by the USRP frontend, as described in Section~\ref{sec:testbench}.

During the coarse synchronization, the results for the PSS detection is shown in Fig.~\ref{fig:PSSDetect}, where the maximum value  $\hat{\epsilon}_{\text{max}}$ of auto-correlation is located at 99230 in samples, and the identification number of PSS is hence $m_2 = 2$. Fig.~\ref{fig:SSSDetect} shows the correlation results of SSS, from which the identification number of SSS is detected as $m_1 = 200$ according to the position of the PSS and the SSS symbols in a SSB. Therefore, the CID can be estimated as $N_{ID}^{cell} = 602$ according to (\ref{eq_pci}). 

\begin{figure}[t]
	\centering
	\begin{minipage}[t]{.5\textwidth}
		\centering
		\includegraphics[width=.85\linewidth]{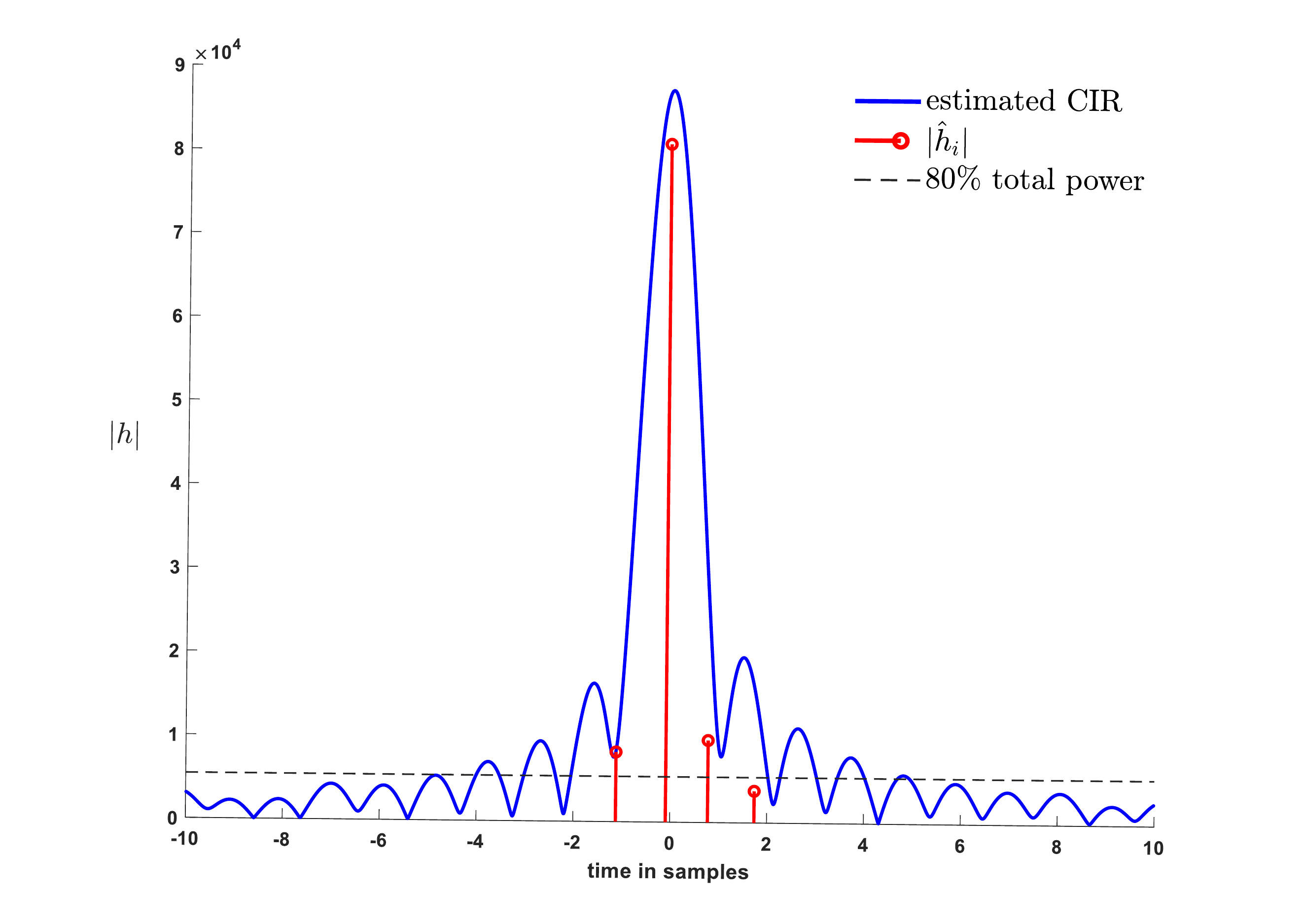}
		\caption{Channel acquisition results in field test.}
		\label{fig:channelEst}
	\end{minipage}
    \\
	\begin{minipage}[t]{.5\textwidth}
		\centering
		\includegraphics[width=\linewidth]{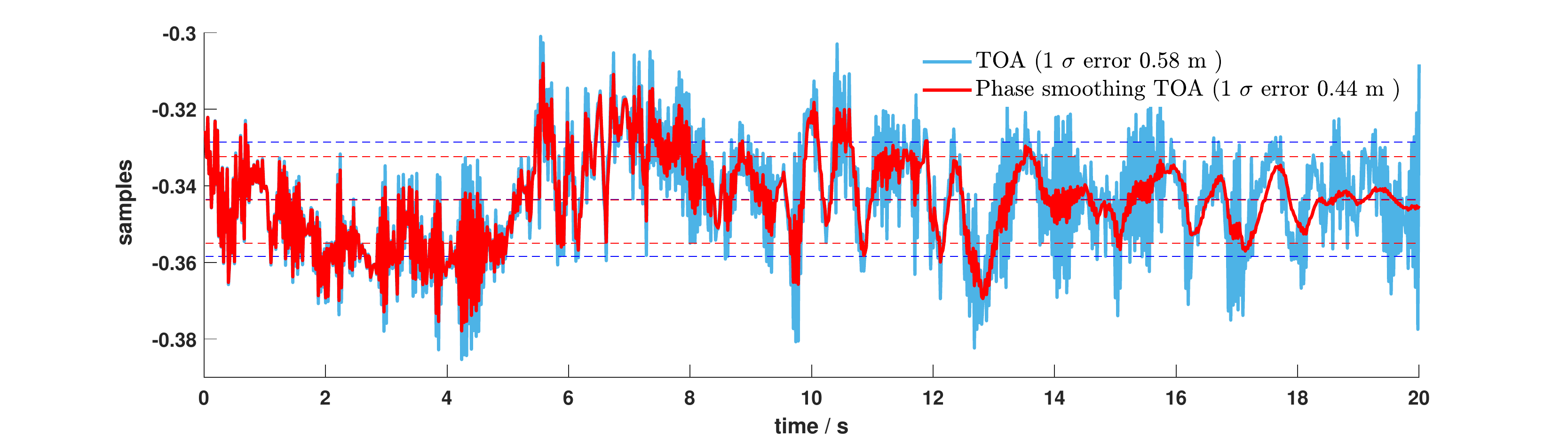}
		\caption{Static test results in field test.}
		\label{fig:staticTest}
	\end{minipage}
\end{figure}

During multipath acquisition, the threshold is set as $80\%$ of the total power within the acquisition region of $D_\rho$. The results given by the multipath acquisition are presented in Fig.~\ref{fig:channelEst}, where four paths are acquired and the corresponding time delays of the paths can be estimated according to section~\ref{sec:multipath_acquisition}.

By setting the detected time delay of the first path as the initial delay, a 20-second DLL tracking process is carried out and the results are shown in  Fig.~\ref{fig:staticTest}. From the tracking results, the 1-$\sigma$ accuracy of the ToA estimation is about 0.58 m. Furthermore, with the assistance of the carrier phases as described in Section~\ref{sec:carrier_phase_estimation}, the ToA tracking results can be further smoothed~\cite{cite34}, yielding the 1-$\sigma$ accuracy of about 0.44 m.
\begin{table}[t]
	\renewcommand\arraystretch{1.4}
	\begin{center}
		\caption{The relevant parameters of the gNB and hardware platform used in field tests}
		\label{table:tests_parameters}
		\resizebox{72mm}{25mm}{
			\begin{tabular}{|c|c|c|c|} 
				\hline
				\multicolumn{3}{|c|}{Parameter}                         & Value \\ \hline
				\multirow{7}{*}{gNB information}    & \multicolumn{2}{c|}{Cellular provider} & China Mobile\\ \cline{2-4}
				& \multicolumn{2}{c|}{Band}              & n41 \\ \cline{2-4}
				& \multicolumn{2}{c|}{SSB pattern}       & Case C \\ \cline{2-4} 
				& \multicolumn{2}{c|}{Network mode}      & NSA \\ \cline{2-4}
				& \multicolumn{2}{c|}{Duplexing}        & TDD \\ \cline{2-4}
				& \multicolumn{2}{c|}{Frequency range}   & 2515-2675 MHz \\ 
				\cline{2-4}
				& \multicolumn{2}{c|}{Center Frequency}  & 2565 MHz
				\\
				\hline                                    				
				\multirow{7}{*}{Receiver information} & \multicolumn{2}{c|}{Hardware type}      & USRP B210\\ \cline{2-4}
				& \multicolumn{2}{c|}{Sampling bandwidth} & 10 MHz \\ \cline{2-4}
				& \multirow{2}{*}{Acquisition} & Threshold  &  80\% \\ \cline{3-4}
				&  & Time interval & 0.1 s \\ \cline{2-4}
				& \multirow{3}{*}{Tracking loop} & Updating time & 20 ms \\ \cline{3-4}
				&  & \multirow{2}{*}{Bandwidth} & 25 Hz (Static Tests)\\ \cline{4-4}
				&  &   & 0.5 Hz (Dynamic Tests)\\
				\hline
			\end{tabular}
		}
	\end{center}
\end{table}

\begin{figure}[!h]
	\begin{minipage}{0.5\textwidth}  
		\centerline{\includegraphics[width=1\textwidth,height=0.11\textheight]{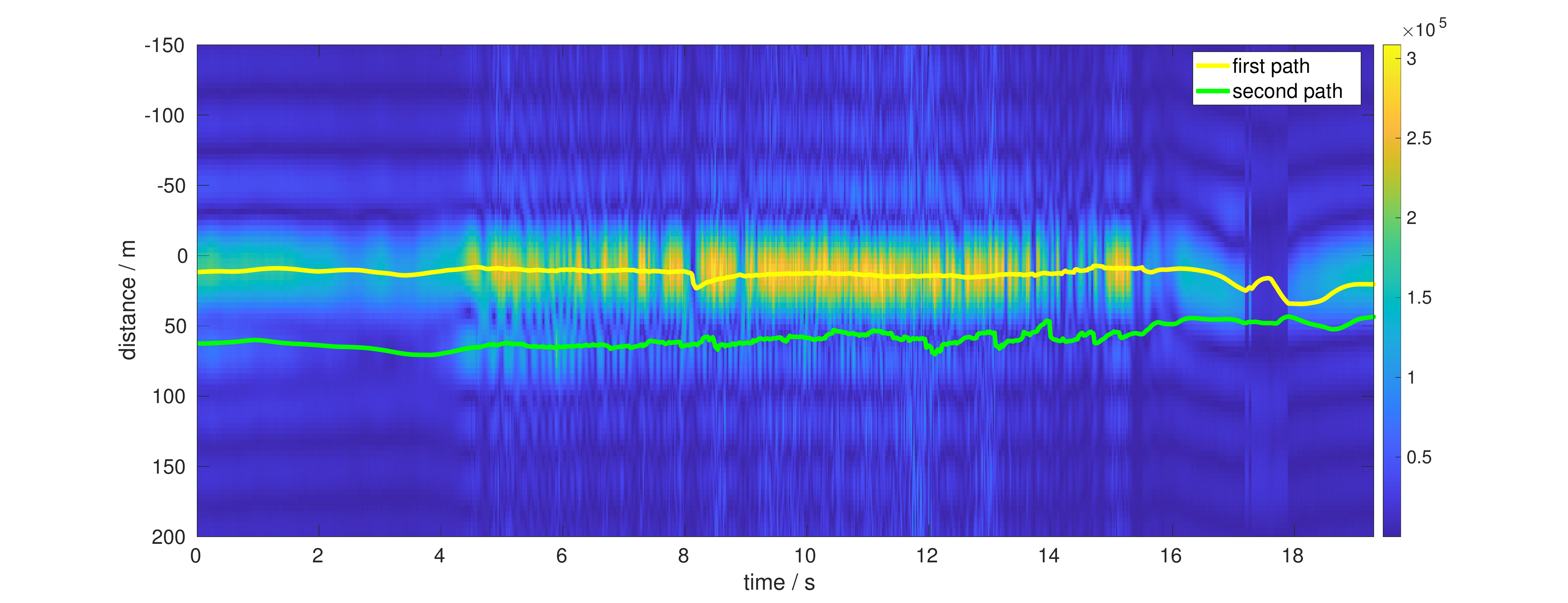}}
		\centerline{(a)}
	\end{minipage}
	\vfill
	\begin{minipage}{0.5\textwidth}  
		\centerline{\includegraphics[width=1\textwidth,height=0.11\textheight]{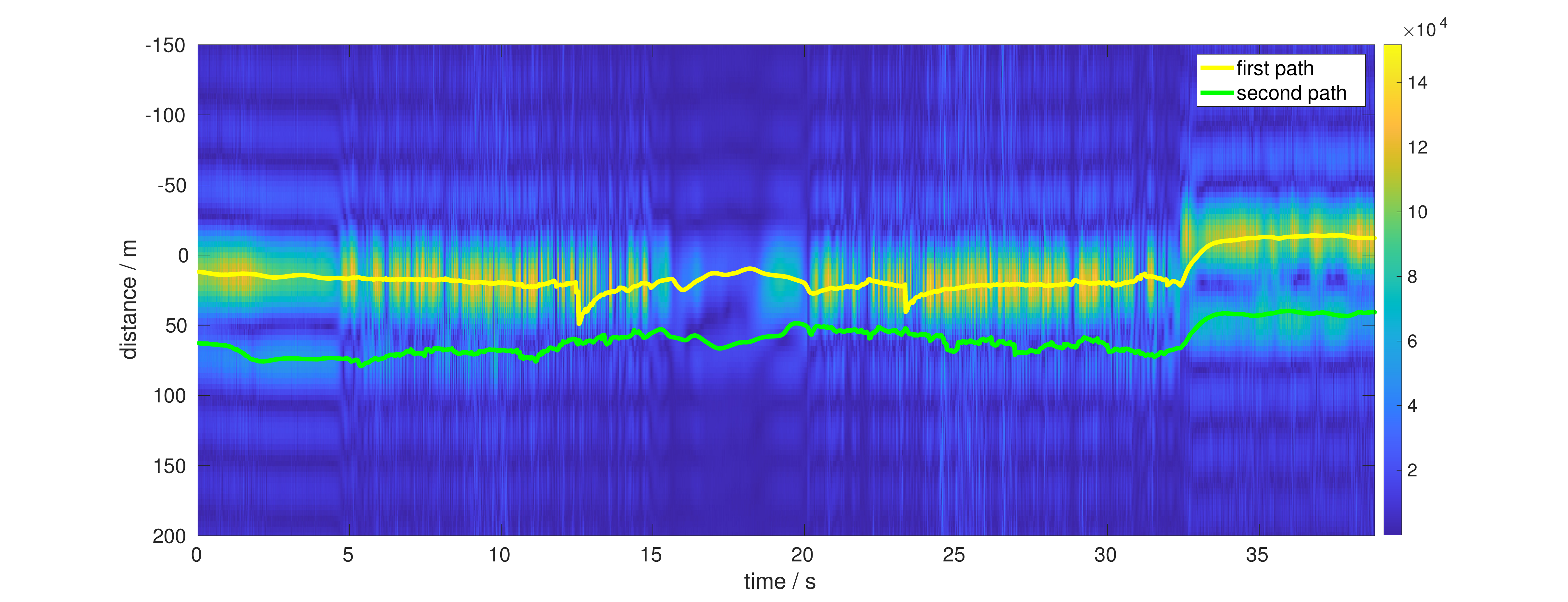}}
		\centerline{(b)}
	\end{minipage}
	\caption{Mutipath tracking results of (a) test 1 and (b) test 3.}
	\label{fig:multiscan}
\end{figure}

\begin{figure}[t]
	\subfigure[]{
		\includegraphics[width=4cm]{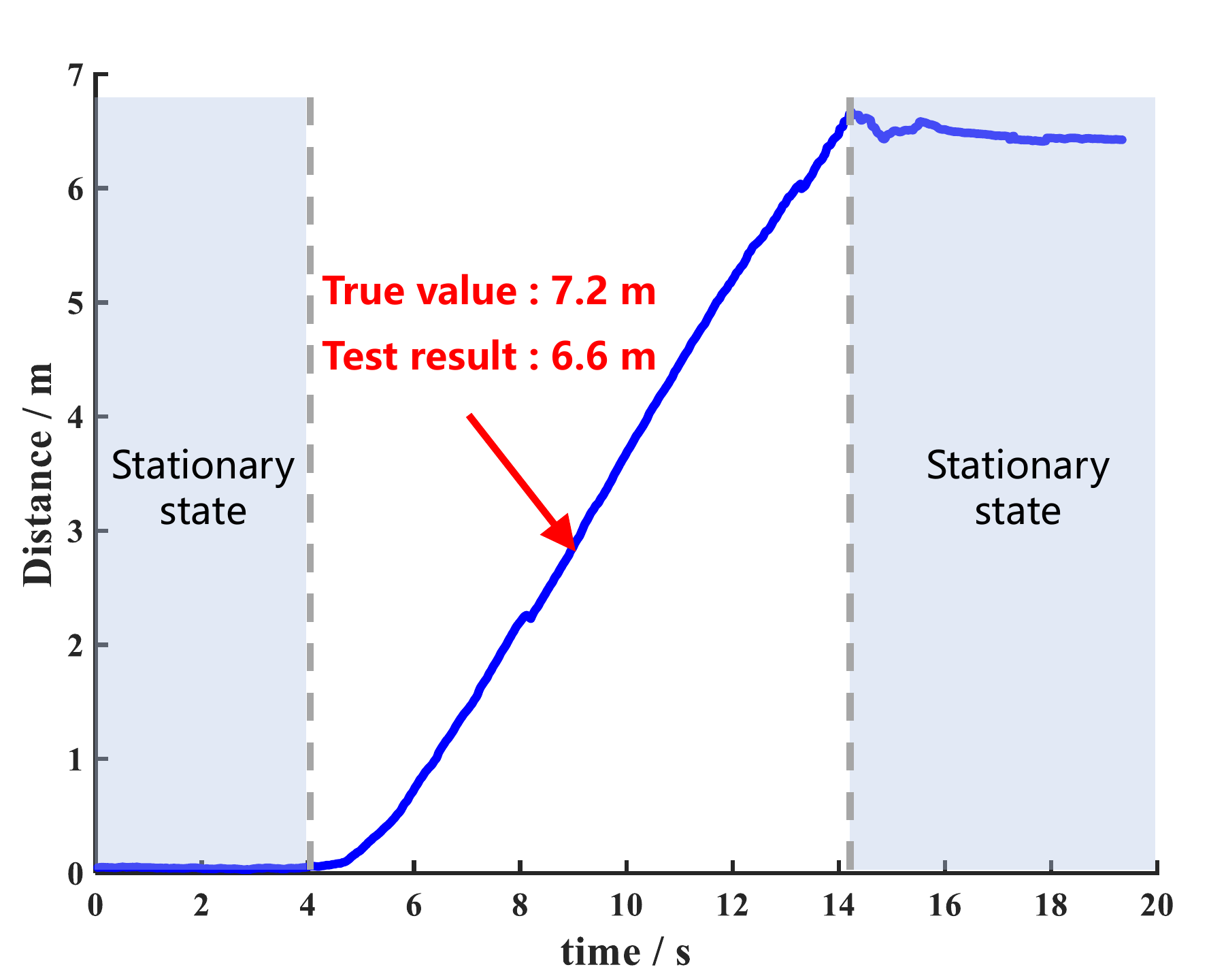}
	}
	\quad
	\subfigure[]{
		\includegraphics[width=4cm]{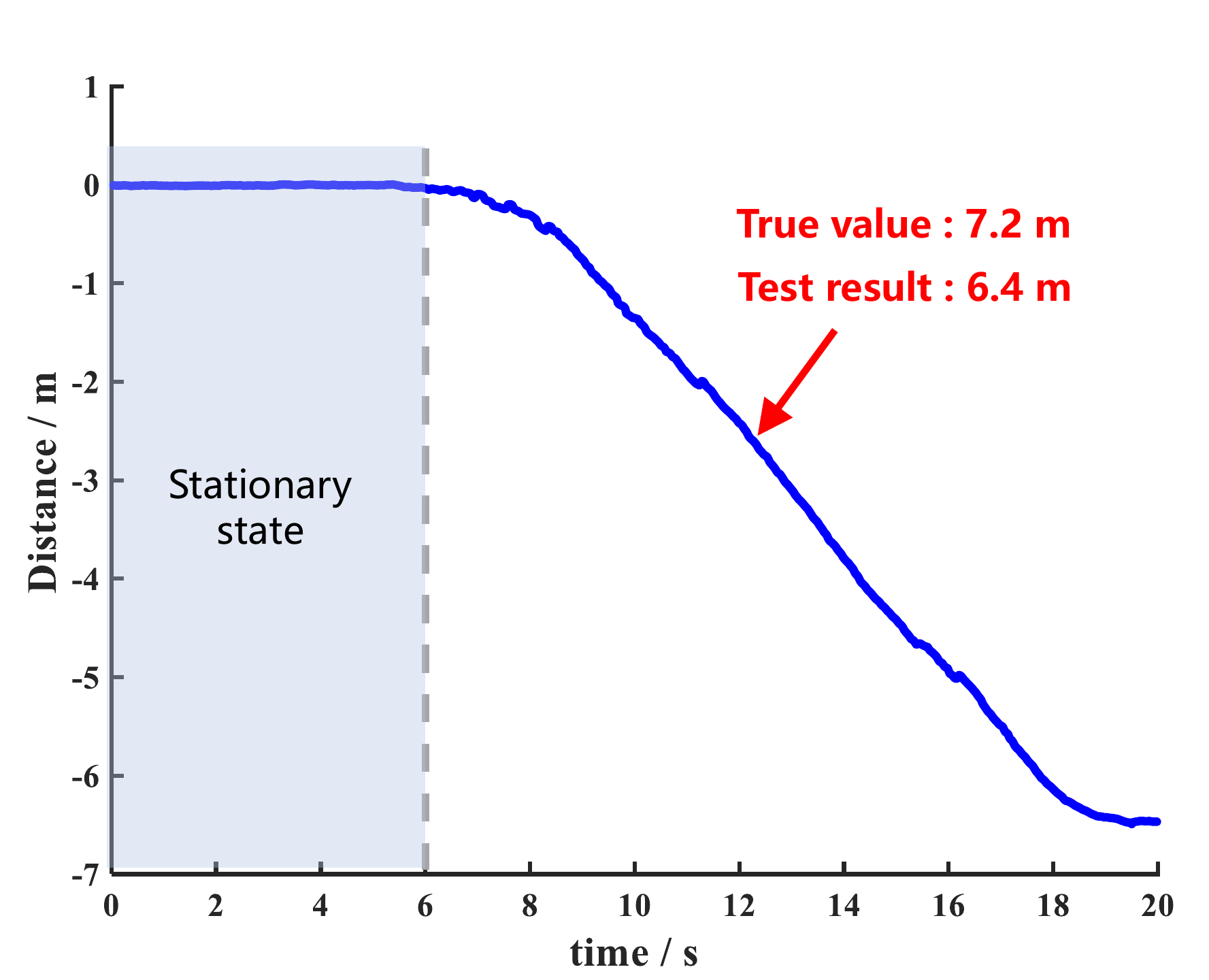}
	}
	\quad
	\subfigure[]{
		\includegraphics[width=4cm]{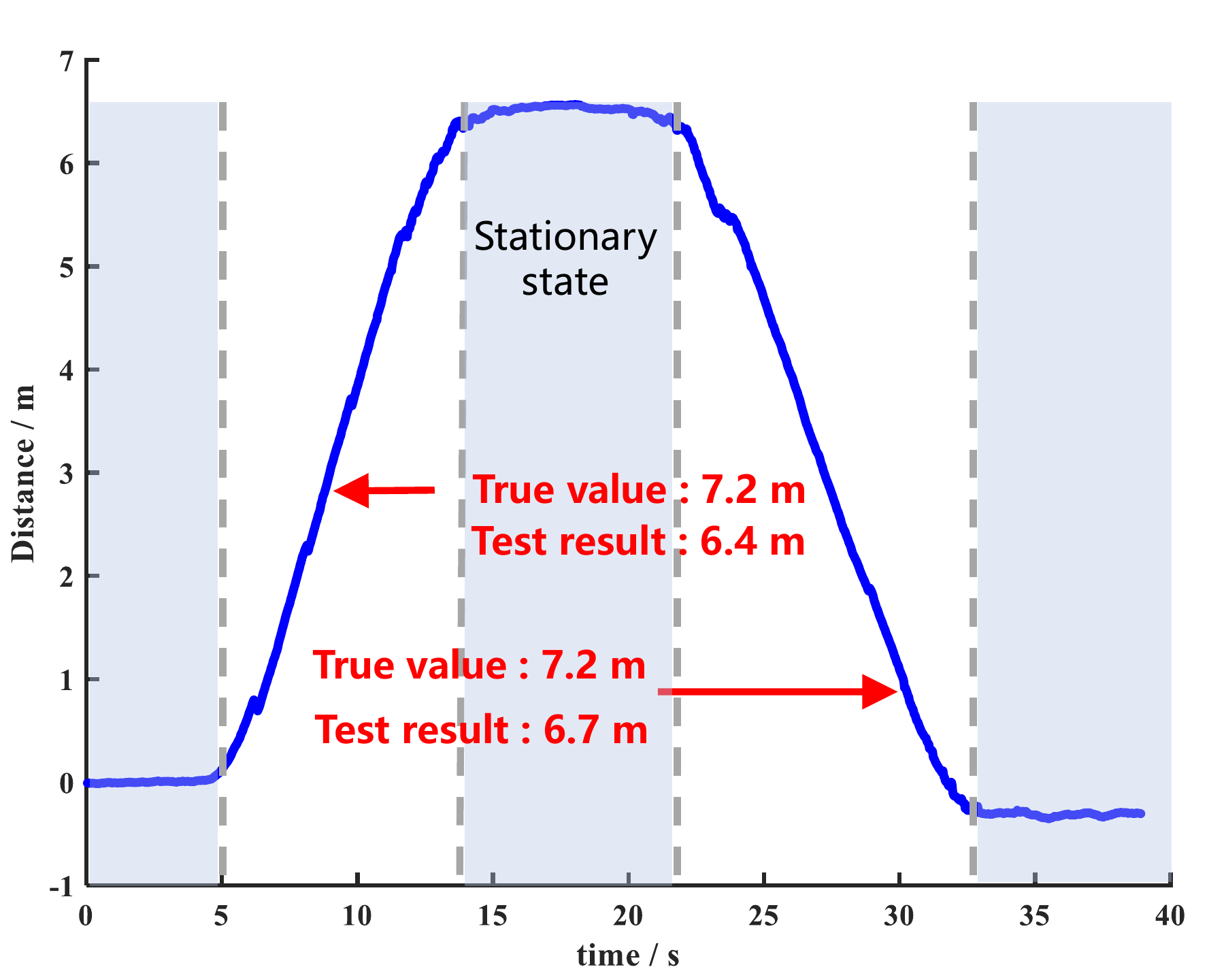}
	}
	\quad
	\subfigure[]{
		\includegraphics[width=4cm]{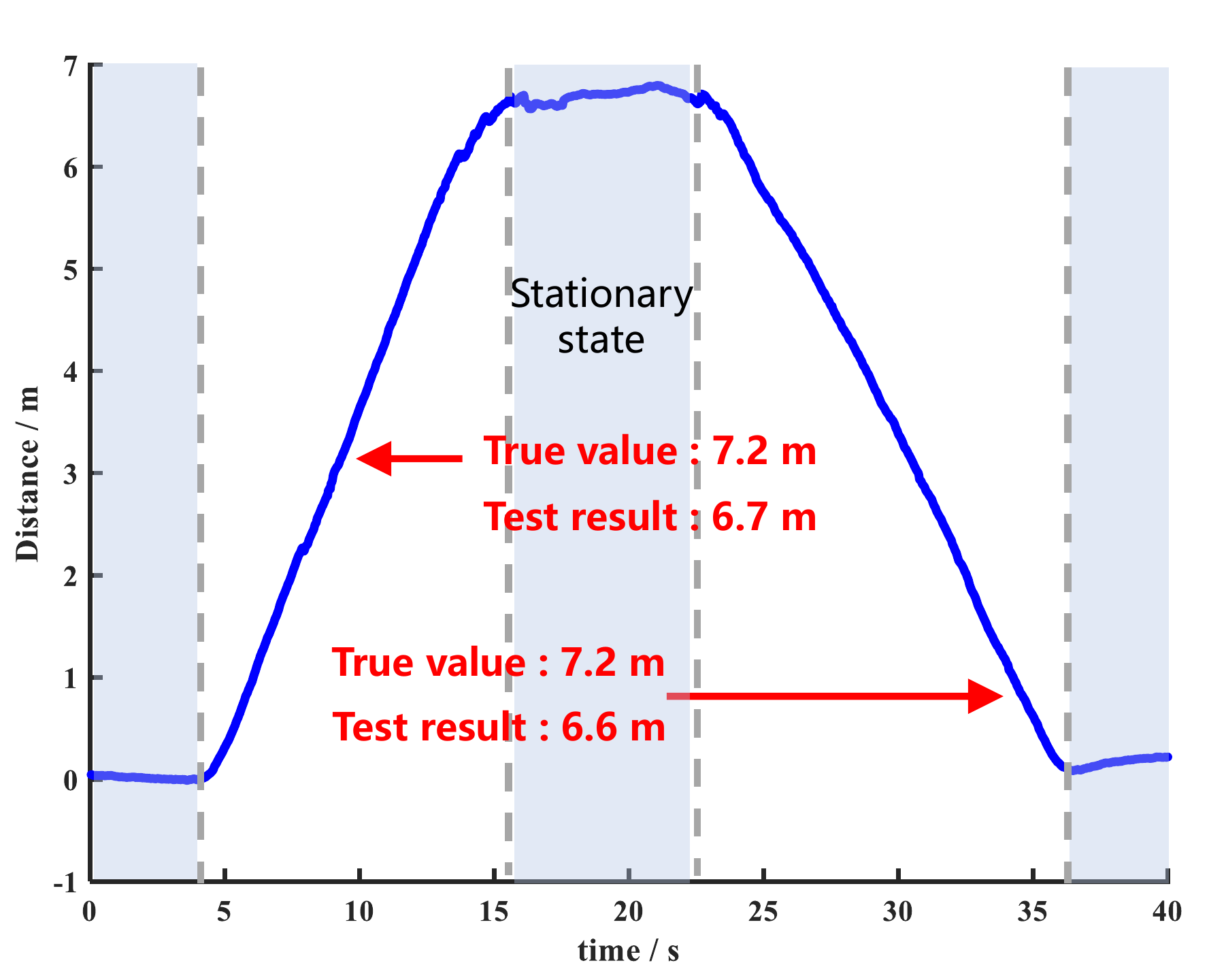}
	}
	\quad
	\subfigure[]{
		\includegraphics[width=4cm]{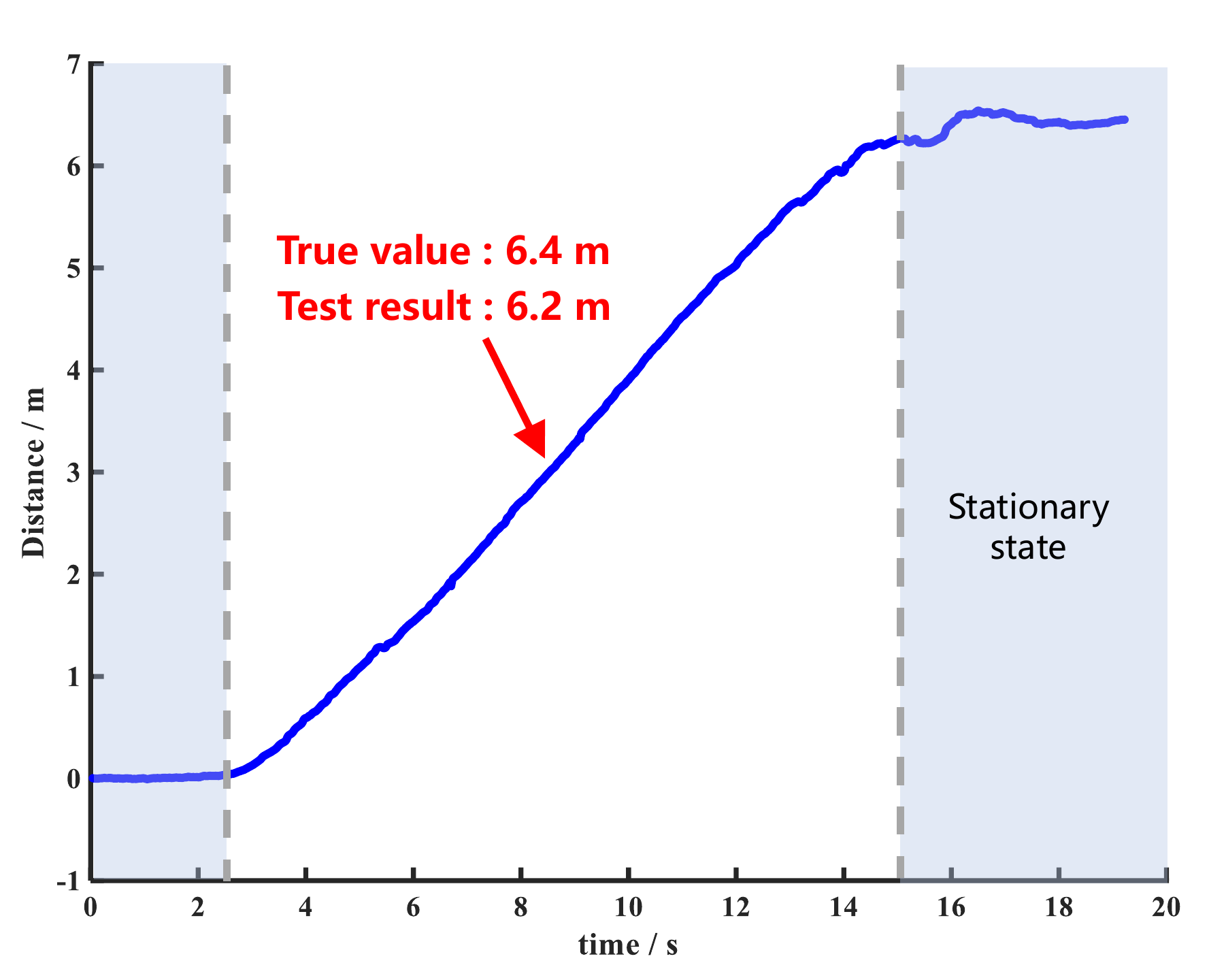} 
	}
	\quad
	\subfigure[]{
		\includegraphics[width=4cm]{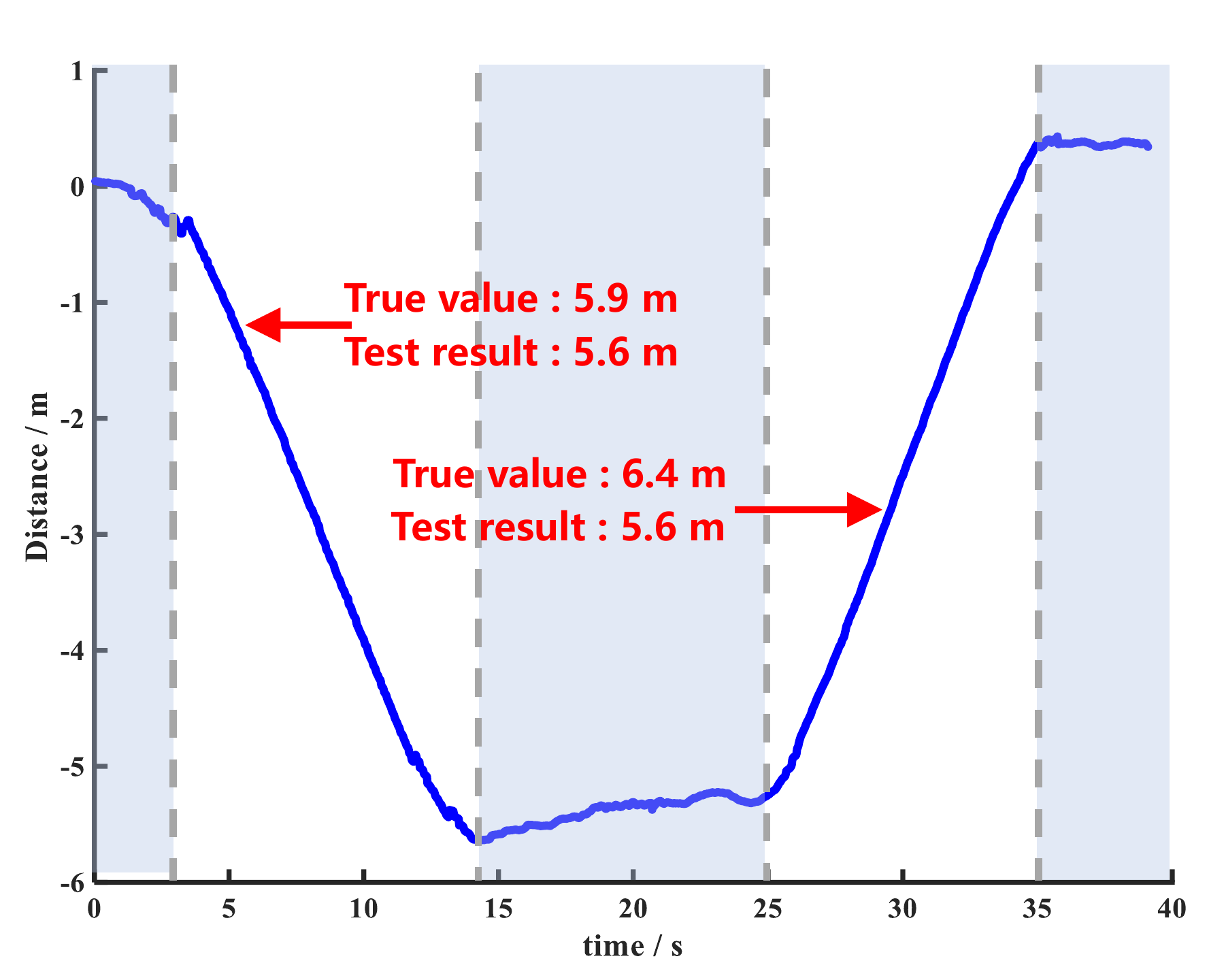} 
	}	
	\caption{Carrier phase ranging based estimation from field tests. (a) test 1. (b) test 2. (c) test 3. (d) test 4. (e) test 5. (f) test 6.}
	\label{fig:dist}
\end{figure}
\subsection{Dynamic Tests on Indoor Pedestrian Walking}
The trajectories of the pedestrian walking for indoor field tests are shown in Fig.~\ref{fig:scenario}. The true moving distance is summarized in Table~\ref{table:dynamic_walking_path}. 
In our tests, a pedestrian walks with a constant speed. At the beginning of each test, the pedestrian stands still for about 5 seconds so as to give enough time to initialize the system. The second column of Table~\ref{table:dynamic_walking_path} shows the results of walking paths tested. Note that in test 1 to test 4, the true distance between point A and point B is 9 m, and the true distance moved is 7.2 m.
In test 5 and test 6, different walking routes were carried out as shown in Fig.~\ref{fig:scenario}.

Fig.~\ref{fig:multiscan} (a) shows the multipath tracking of the incoming signals of test 1, when the pedestrian walks from B to A. Fig.~\ref{fig:multiscan} (b) shows the multipath tracking results of test 3 when the pedestrian first walks from B to A and then returns to B. In both tests, two paths are detected, which carry $80\%$ of the whole multipath channel. In (a) and (b) of Fig.~\ref{fig:multiscan}, the heat maps correspond to the magnitudes of CIR at the corresponding sampling times. As seen from Fig.~\ref{fig:multiscan}, the multipaths are reliably detected and tracked, which are in accordance with the highlights of the heat map.

\begin{figure}[!h]
	\begin{minipage}{.5\textwidth}  
		\centerline{\includegraphics[width=.85\textwidth]{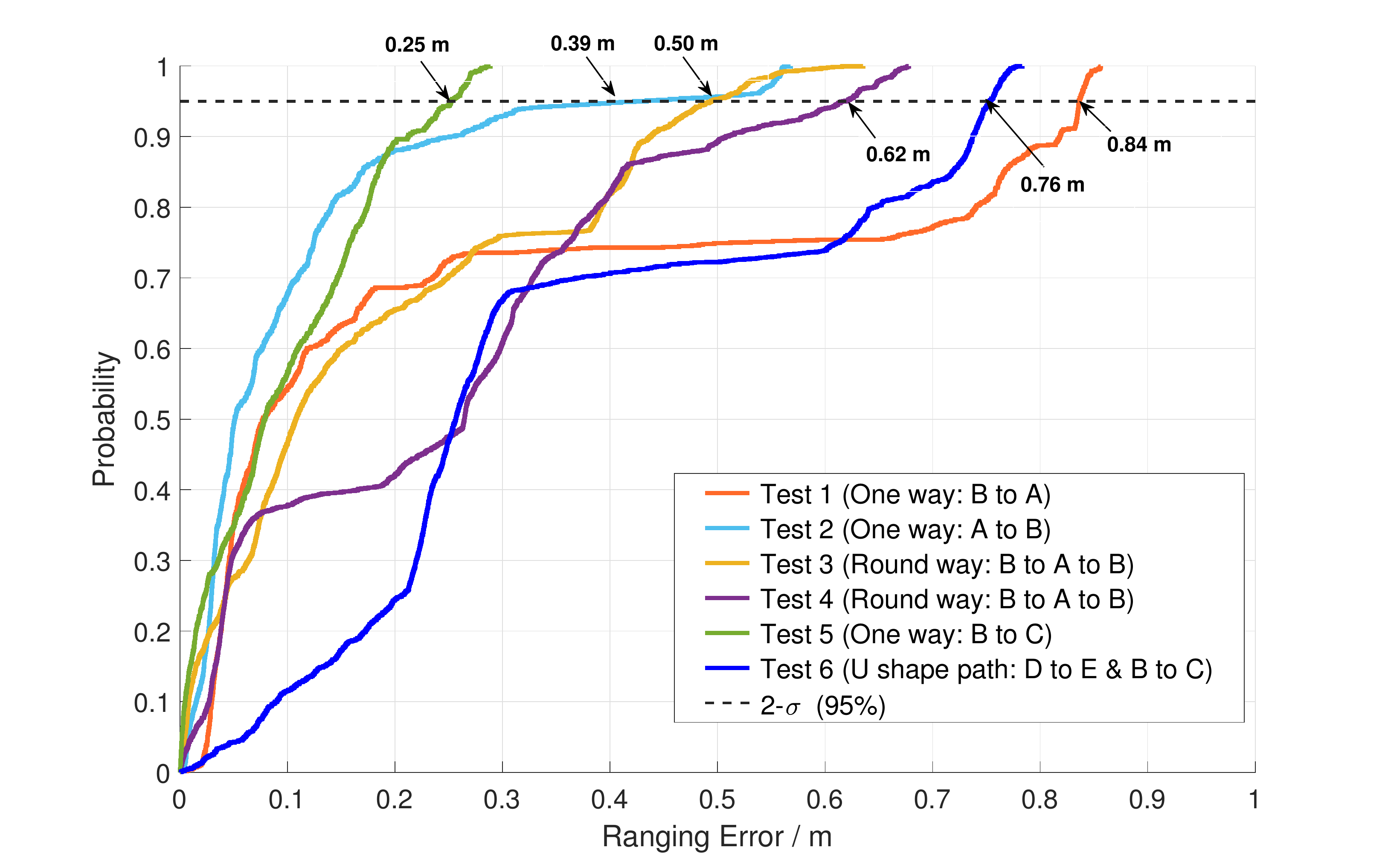}}
		\centerline{(a)}
	\end{minipage}
	\hfill
	\begin{minipage}{.5\textwidth}  
		\centerline{\includegraphics[width=.85\textwidth]{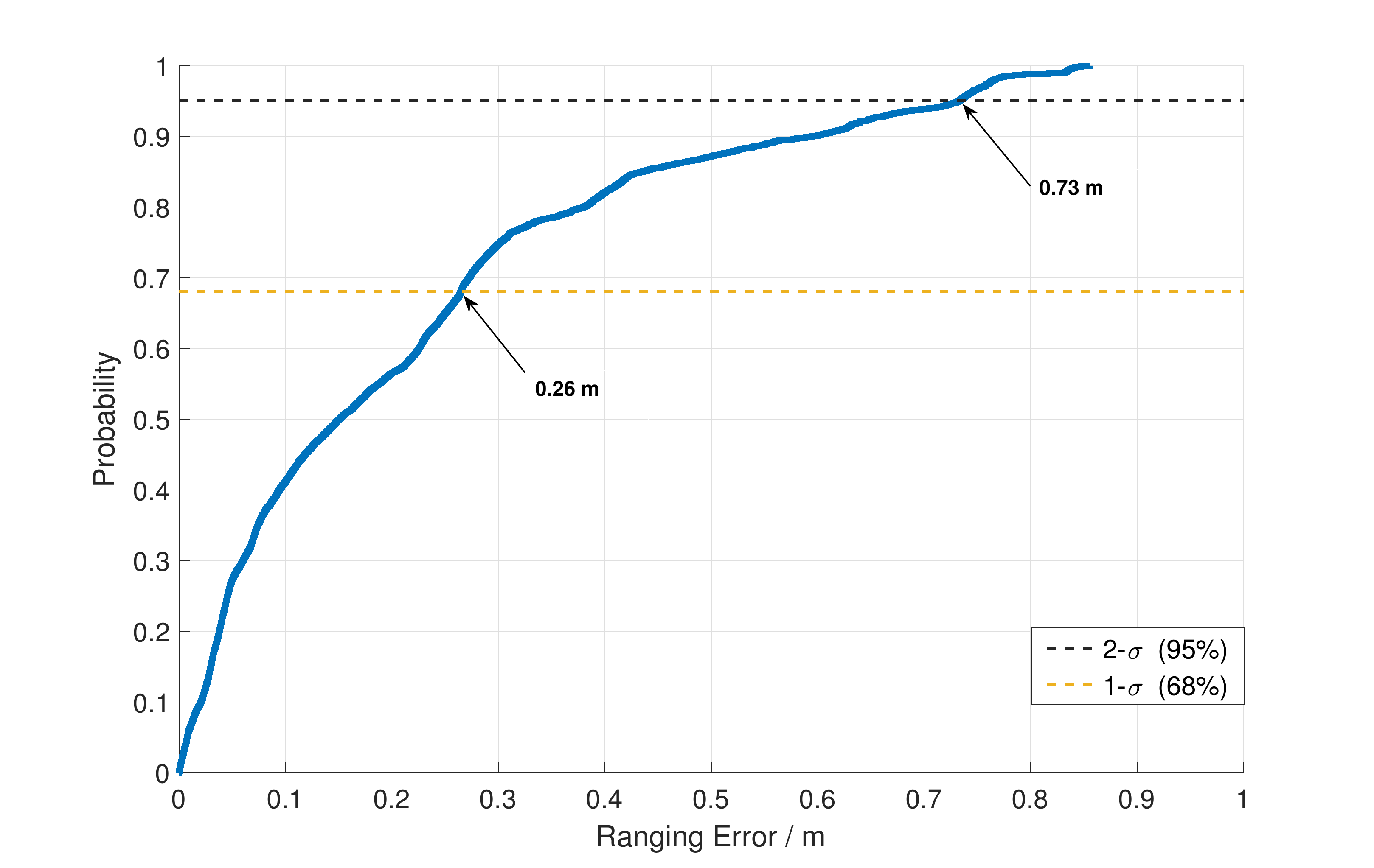}}
		\centerline{(b)}
	\end{minipage}
	\caption{(a) Comparison of the ranging errors in different dynmaic tests. (b) CDF of the ranging errors of the dynamic tests.}
	\label{fig:DistCDF}
\end{figure}

Fig.~\ref{fig:dist} shows the estimation of the pedestrian's walking distance obtained by the carrier phase ranging method. For these figures, the distance of the starting point is 0 m. When the test pedestrian walks towards the base station, the ranging result is positive, while the ranging distance is negative \textit{vice versa}. The estimation errors of the total walking distance for the tests are listed in the last column of Table~\ref{table:dynamic_walking_path}, which are less than 0.8 m. 
Correspondingly, in Fig.~\ref{fig:DistCDF}, the empirical cumulated distribution functions (CDFs) of the ranging errors obtained from the different tests are demonstrated. It can be shown that for each of the tests, the probability that the estimation error is smaller than 0.9 m at 95\% ( 2-$\sigma$), and the estimation error is smaller than 0.3 m at 1-$\sigma$ accuracy as well as 0.8 m at 2-$\sigma$ accuracy for all dynamic tests.

\begin{table}[t]
	\renewcommand\arraystretch{1.5}
	\centering
	\begin{center}
		\caption{Results of carrier phase assisted ranging estimation} \label{table:dynamic_walking_path}
		\resizebox{90mm}{22mm}{
			\begin{tabular}{|c|c|c|c|c|c|} 
				\hline
				\multirow{2}{*}{Test} &
				\multicolumn{2}{c|}{\multirow{2}{*}{Moving routes}} & Distance & Real moving & \multirow{2}{*}{Distance error (m)}\\
				 & \multicolumn{2}{c|}{} & measured (m) & distance (m) & \\
				\hline
				1 & One way & B $\rightarrow$ A & 6.6 & \multirow{6}{*}{7.2} & 0.6\\
				\cline{1-4}
				\cline{6-6}
				2 & One way & A $\rightarrow$ B & 6.4 & & 0.8\\
				\cline{1-4}
				\cline{6-6}
				\multirow{2}{*}{3} & \multirow{2}{*}{Round way} & B $\rightarrow$ A & 6.4 & & 0.8\\
				\cline{3-4} \cline{6-6}
				&  & A $\rightarrow$ B & 6.7 & & 0.5\\
				\cline{1-4} \cline{6-6}
				\multirow{2}{*}{4} & \multirow{2}{*}{Round way} & B $\rightarrow$ A & 6.7 & & 0.5\\
				\cline{3-4} \cline{6-6}
				&  & A $\rightarrow$ B & 6.6 & & 0.6\\
				\cline{1-4}
				\cline{6-6}
				\hline
				5 & One way & B $\rightarrow$ C & 6.2 & 6.4 & 0.2\\
				\hline
				\multirow{2}{*}{6} & \multirow{2}{*}{U shape path} & D $\rightarrow$ E & 5.6 & 5.9 & 0.3\\
				\cline{3-6}
				& & B $\rightarrow$ C & 5.6 & 6.4 & 0.8\\
				\hline
			\end{tabular}
		}
	\end{center}
\end{table}
\section{Conclusion}~\label{sec:conclusion}
Since  the 5G cellular networks were started to be deployed worldwide, the positioning based on 5G NR signals has drawn a lot of attention. In this paper, the problem of ToA estimation for indoor positioning in 5G NR environment was studied. A USRP based SDR receiver was developed, which includes coarse synchronization, multipath acquisition, multipath tracking and the carrier phase ranging estimation.

To evaluate the accuracy of the proposed method, indoor field tests were carried out in an office area, where a 5G NR gNB is deployed for commercial use. The test results show that, in the static scenarios, the ToA accuracy is about 0.5 m, while in the pedestrian mobile scenarios, the range accuracy with 2-$\sigma$ error is within 0.8 m. Theoretical analysis was also carried out, showing that, in comparison with GPS L1 C/A and GAL E1, the DM-RS in the SSB of the 5G NR has a narrower main lobe of ACF which is promising for navigation and positioning.

Our future research will consider the positioning with the triangulation method in the indoor communications environment where multiple 5G gNBs may be heard. We will also consider the sensor fusion method by combining the carrier phase ranging estimation with the other information obtained from the smartphone embedded sensors, such as the inertial sensors, camera, BLE/WiFi/UWB, etc. In addition, we will also consider the issue of transmitter fusion in 5G NR positioning, which is very common in positioning in signal frequency networks~\cite{SFN}.

\section*{Acknowledgment}
The research is supported by the National Key Research and Development Programs, grant number 2018YFB0505400.


%




\ifCLASSOPTIONcaptionsoff
  \newpage
\fi



%

\bibliographystyle{IEEEtran}

%

%
\begin{IEEEbiography}
[{\includegraphics[width=1in,height=1.25in,clip,keepaspectratio]{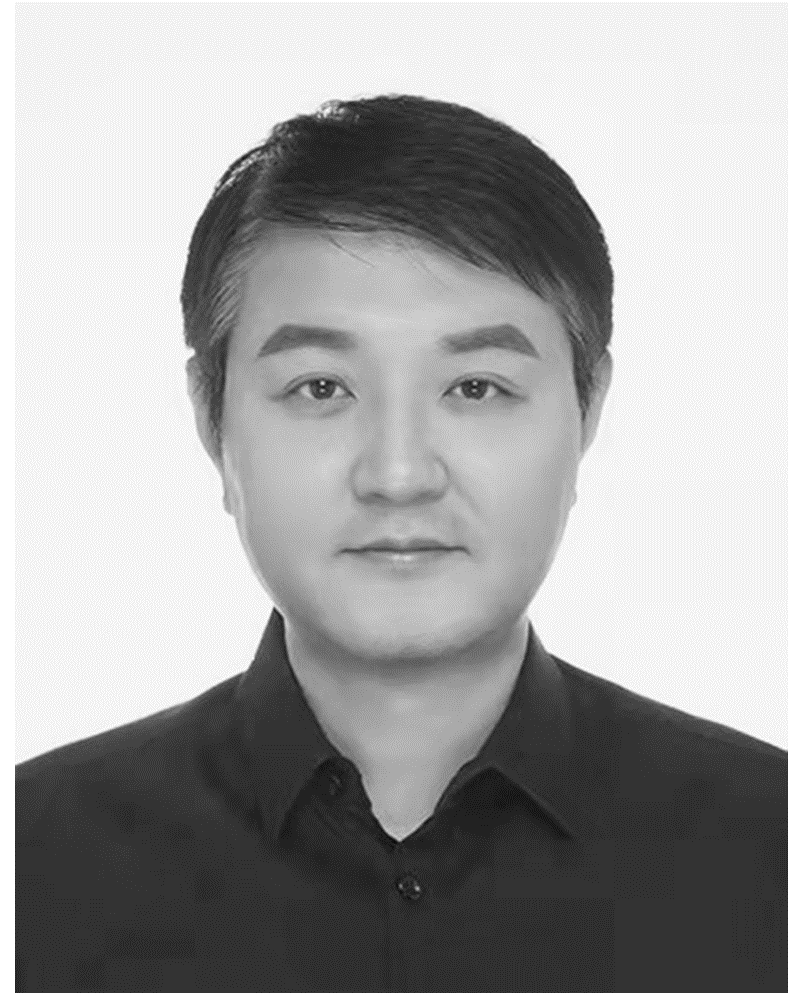}}]{Liang Chen}
received the Ph.D. degree from Southeast University, Nanjing, China, in 2009. From August 2009 to December 2010, he was a Research Associate with Tampere University of Technology, Finland and from January 2011 to February 2017, he was a Senior Research Scientist with Finnish Geospatial Research Institute. In March 2017, he has been with Wuhan University, China, where he is the Professor of the State Key Laboratory of Surveying, Mapping and Remote Sensing Information Engineering, as well as the Deputy Head of Institute of Artificial Intelligence in Geomatics. He has published over 70 scientific research papers. He served as an Associate Editor to Navigation, the journal of Institute of Navigation, Journal of Navigation. His research interests include signal processing for navigation and positioning, indoor positioning.
\end{IEEEbiography}



\begin{IEEEbiography}
[{\includegraphics[width=1in,height=1.25in,clip,keepaspectratio]{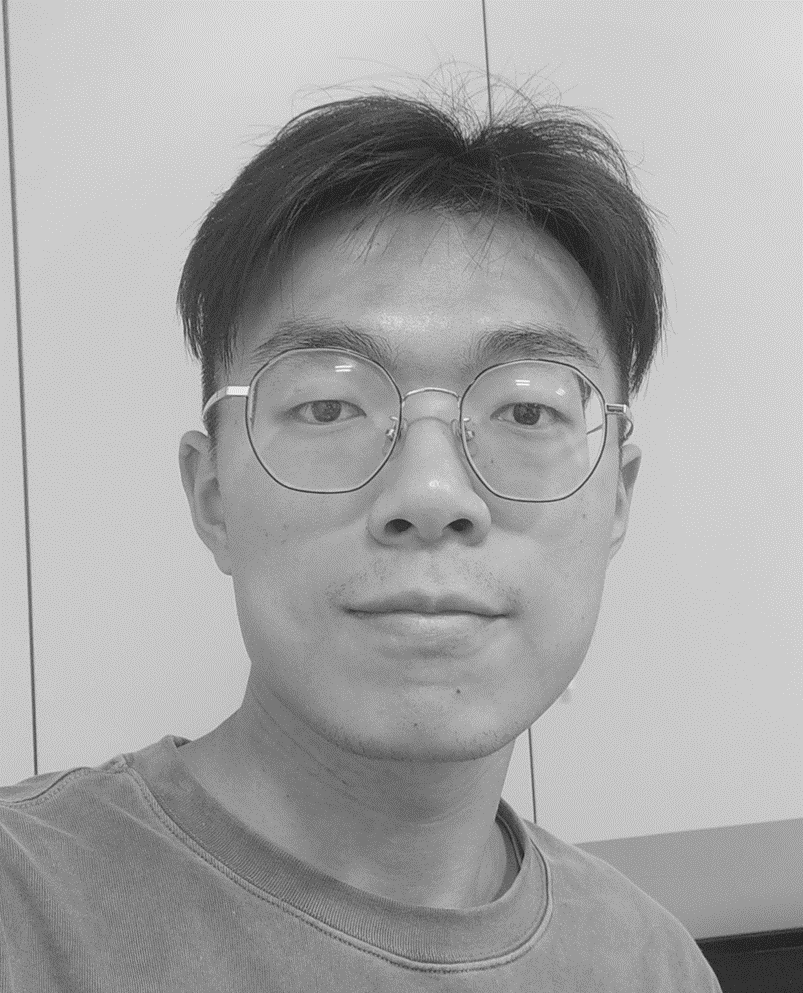}}]{Xin Zhou}
received the B.S. and M.S. degrees from Wuhan University, Wuhan, China, in 2018 and 2020, respectively. He is currently pursuing the Ph.D. degree in geodesy and survey engineering with the State Key Laboratory of Information Engineering in Surveying, Mapping and Remote Sensing, Wuhan University. 

His research interests include indoor positioning and navigation technology based on signal of opportunity, wireless communications, and the Internet of Things.
\end{IEEEbiography}

\begin{IEEEbiography}
[{\includegraphics[width=1in,height=1.25in,clip,keepaspectratio]{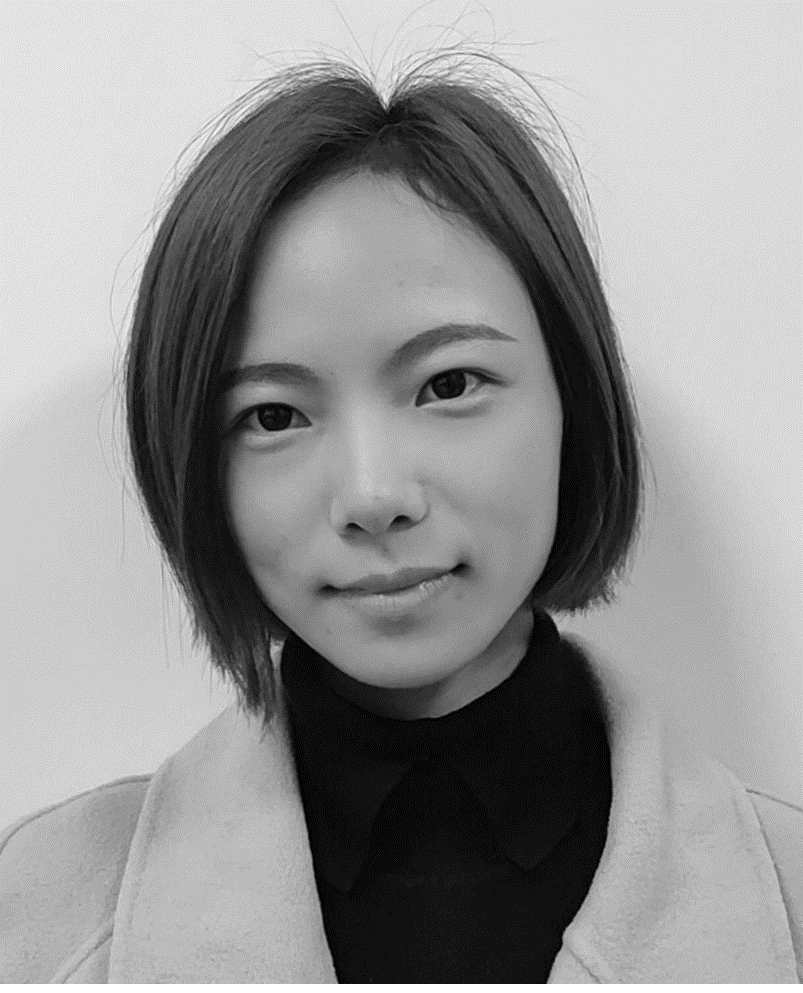}}]{Feifei Chen}
received the B.S. degree from Anhui University of Science and Technology, Huainan, China, in 2014. She is currently pursuing the M.S. degree in geomatics engineering with the State Key Laboratory of Information Engineering in Surveying, Mapping and Remote Sensing, Wuhan University. 

Her research interests include indoor positioning and navigation technology based on signal of opportunity, and wireless communications.
\end{IEEEbiography}

\begin{IEEEbiography}
[{\includegraphics[width=1in,height=1.25in,clip,keepaspectratio]{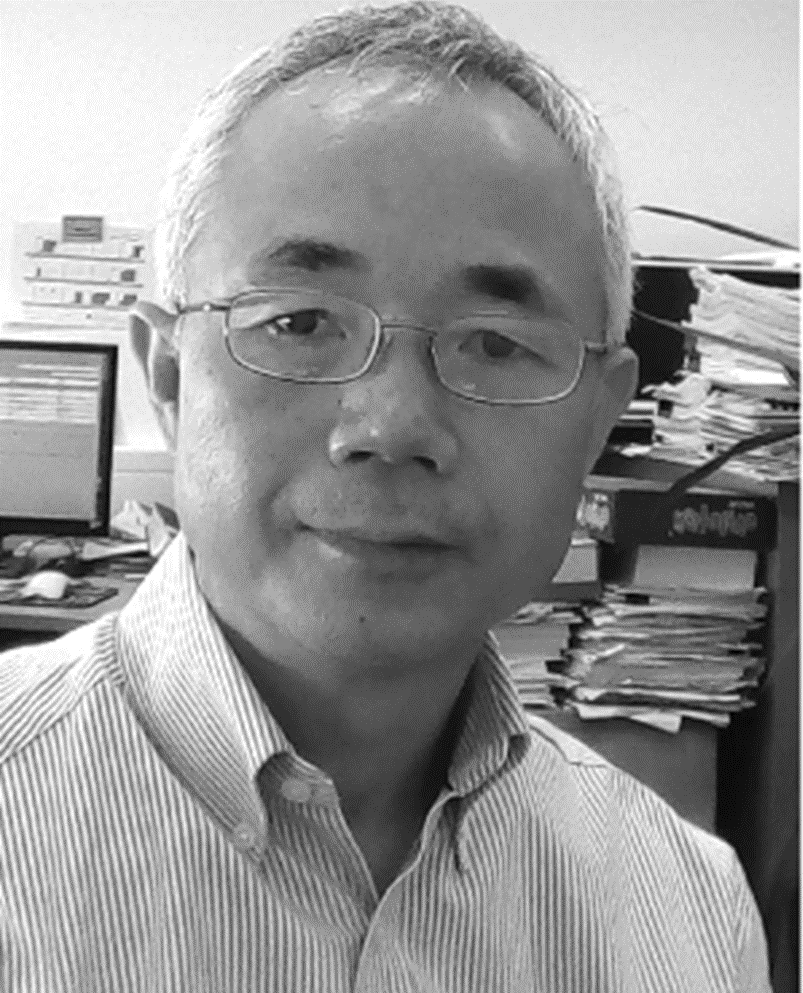}}]{Lie-Liang Yang}
received the B.Eng. degree in communications engineering from Shanghai Tiedao University, Shanghai, China, in 1988, and the M.Eng. and Ph.D. degrees in communications and electronics from Northern (Beijing) Jiaotong University, Beijing, China, in 1991 and 1997, respectively. From June 1997 to December 1997, he was a Visiting Scientist with the Institute of Radio Engineering and Electronics, Academy of Sciences of the Czech Republic. Since December 1997, he has been with the University of Southampton, U.K., where he is the Professor of wireless communications with the School of Electronics and Computer Science. He has research interest in wireless communications, wireless networks and signal processing for wireless communications, as well as molecular communications and nano-networks. He has published over 390 research papers in journals and conference proceedings, authored/coauthored three books, and also published several book chapters. He is a Fellow of the IET and was a Distinguished Lecturer of the IEEE VTS. He served as an Associate Editor to the IEEE TRANSACTIONS ON VEHICULAR TECHNOLOGY and Journal of Communications and Networks (JCN), and is currently an Associate Editor to IEEE ACCESS and a Subject Editor to Electronics Letters.
\end{IEEEbiography}

\begin{IEEEbiography}
[{\includegraphics[width=1in,height=1.25in,clip,keepaspectratio]{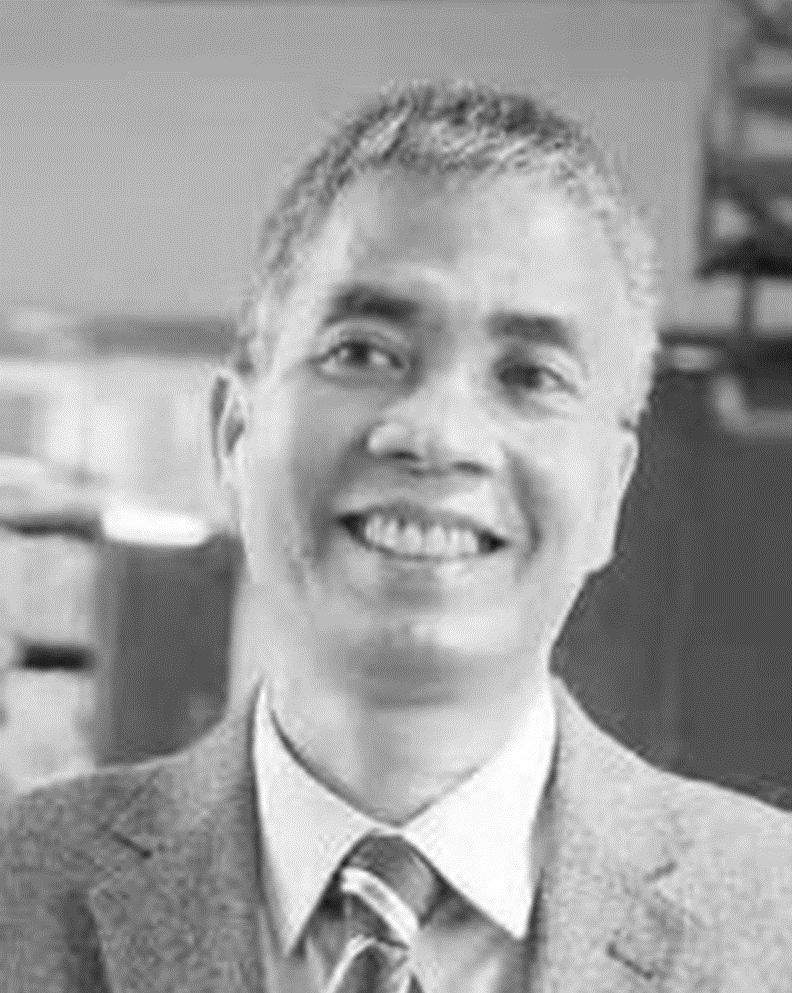}}]{Ruizhi Chen}
received the Ph.D. degree from the University of Helsinki, Helsinki, Finland, in 1991. He is currently a Professor and the Director of the State Key Laboratory of Information Engineering in Surveying, Mapping and Remote sensing, Wuhan University, Wuhan, China. He was an Endowed Chair Professor with Texas A\&M University Corpus Christi, Corpus Christi, TX, USA, the Head and a Professor with the Department of Navigation and Positioning, Finnish Geodetic Institute, Kirkkonummi, Finland, and the Engineering Manager of Nokia, Espoo, Finland. He has published two books and more than 200 scientific papers. His current research interests include indoor positioning, satellite navigation, and location-based services.

\end{IEEEbiography}





\end{document}